\documentclass[12pt, english, twoside]{article}

\usepackage[utf8]{inputenc}
\usepackage[T1]{fontenc}
\usepackage[top=3cm, bottom=3cm, left=2.5cm, right=2.5cm]{geometry}
\usepackage{hyperref}
\hypersetup{pdfborder={0 0 0}, colorlinks=true, breaklinks=true, citecolor=red, linkcolor=blue}
\usepackage{titlesec}
\usepackage{fancyhdr}
\usepackage{graphicx}
\usepackage{graphics}
\usepackage{multirow}
\usepackage{pdfpages}
\usepackage{lmodern}
\usepackage{amsmath,amssymb}
\usepackage{mathrsfs}
\usepackage{url}
\usepackage{caption}
\usepackage{subcaption}
\usepackage{xcolor}
\usepackage[squaren,Gray]{SIunits}

\rmfamily
\DeclareFontShape{T1}{lmr}{b}{sc}{<->ssub*cmr/bx/sc}{}
\DeclareFontShape{T1}{lmr}{bx}{sc}{<->ssub*cmr/bx/sc}{}

\titlespacing*{\section}{0cm}{3ex}{2.5ex}
\titlespacing*{\subsection}{0.5cm}{2.5ex}{2ex}

\begin{document}

% Mise en page

\setlength{\parindent}{0.5cm}
\setlength\abovecaptionskip{-0.3cm}
\setlength\belowcaptionskip{-0.2cm}
\renewcommand{\headheight}{15.2pt}
%\renewcommand{\thesection}{\arabic{section}.\hspace*{-0.2cm}}
%\renewcommand{\thesubsection}{\arabic{section}.\arabic{subsection}.\hspace*{-0.2cm}}
%\renewcommand{\bibname}{References}

% Page de garde

\begin{minipage}[c]{0.32\linewidth}
\center
\includegraphics[width=50mm]{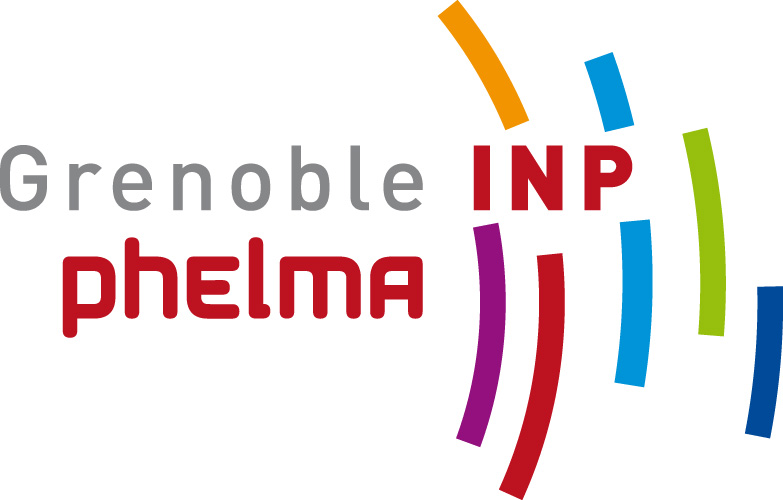}
\end{minipage}
\hfill
\begin{minipage}[c]{0.35\linewidth}
\center
\includegraphics[width=45mm]{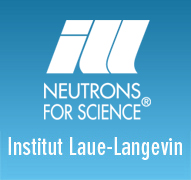}
\end{minipage}

\vspace{1.3cm}

\begin{center}
\normalsize \textsc{Grenoble INP - Phelma} \\
\normalsize \textsc{Ecole Nationale Superieure de Physique, Electronique et Mat\'eriaux} \\
\normalsize Minatec - 3 parvis Louis N\'eel BP 257 \\
\normalsize 38016 Grenoble Cedex 1 \\
~\\
\normalsize \textsc{Master 2 Recherche-Energetique Physique} \\
\normalsize spe. \textsc{Energie Nucleaire} \\
\end{center}

\vspace{0.2cm}

\begin{center} \fontfamily{ptm} \Large \textit{Internship Report} \end{center}
\vspace{-0.3cm} \rule{\linewidth}{1pt} \vspace{-0.7cm}
\begin{center} \fontfamily{ptm} \Large \textbf{Characterization of the new Data Acquisition System and the Detector Alignment of the \textbf{a}SPECT experiment} \end{center}
\vspace{-0.2cm} \rule{\linewidth}{1pt}

\vspace{0.1cm}

\begin{center} 
\Large \textbf{Romain \textsc{Virot}} \\
~\\
\normalsize26/08/2013
~\\
~\\
\normalsize Directeur de stage : Mr Torsten \textsc{Soldner} \\
~\\
\normalsize President de jury : Mr Gregoire \textsc{Kessedjian} \\
\normalsize Tuteur Phelma : Mr Christophe \textsc{Sage} \\
\normalsize Responsable du Master : Mme Elsa \textsc{Merle-Lucotte} \\
~\\
~\\
\normalsize \textsc{Institut Max Von Laue - Paul Langevin (ILL)} \\
\normalsize \textsc{Nuclear and Particle Physics Group} \\
\normalsize 6, rue Jules Horowitz - BP 156 - 38042 Grenoble Cedex 9 - France \\
\end{center}

\thispagestyle{empty}

% Table des matières et table des figures

\newpage
\pagenumbering{roman}
\setcounter{page}{1}
\setlength{\baselineskip}{0.2in}
\tableofcontents
\thispagestyle{plain}

\newpage
\listoffigures
\vspace{8.5cm}
\thispagestyle{plain}

% Début du rapport

\newpage
\pagenumbering{arabic}
\setcounter{page}{1}

%%%%%%%%%%%%%%%%%%%%%%%%%%%%%%%%%%%%%%%%%%%%%%%%%%%%%%%%%%%%%%%%%%%%%%%%%%%%%%%
%  Introduction
%%%%%%%%%%%%%%%%%%%%%%%%%%%%%%%%%%%%%%%%%%%%%%%%%%%%%%%%%%%%%%%%%%%%%%%%%%%%%%%

\section*{Introduction} \addcontentsline{toc}{section}{Introduction}

The Standard Model (SM) of particle physics is a theory born in the 1960s from the combination of the electromagnetic and weak interactions realized by Sheldon Lee \textsc{Glashow}. After several evolutions, directed by both theoretical and experimental researches, the current formulation was finalized in the mid 1970s, resulting in a theory that describes the electromagnetic, weak, and strong interactions. Even though the SM explains a wide variety of phenomenons it is not perfect: there are several observations that are not explained or compatible with the SM, as for example the matter/antimatter asymmetry or the gravitational interaction. Those implies that it might be a ``low-energy'' limit of a more general model. In order to probe such limits and to test new theories beyond the SM, like the SUper SYmmetry for example, several parameters can be studied. The $a$SPECT experiment is mainly dedicated to the measurement, with a very high precision, of one of them: the electron antineutrino angular correlation coefficient $a$. As shown in the section \ref{theorie} \,, this coefficient can be derived from the proton recoil spectrum in free neutron beta decay ($\beta^{-}$).

From April 2013 to the beginning of August 2013, the aSPECT spectrometer was installed on a cold neutron beam facility at the Institut Laue-Langevin (ILL) in order to realize a high-precision measurement of a. Among all the upgrades that were made during the past years, a new state-of-the-art acquisition system has been tested during this beam time.

This report will briefly introduce the theory behind the $a$SPECT concept. We will then present the experiment itself and focus on the detector and the electronic processing of the signals. After a study of the beam and detector position, we will present the new acquisition system in detail and take a closer look to his behavior.

%%%%%%%%%%%%%%%%%%%%%%%%%%%%%%%%%%%%%%%%%%%%%%%%%%%%%%%%%%%%%%%%%%%%%%%%%%%%%%%
%  Présentation de l'ILL
%%%%%%%%%%%%%%%%%%%%%%%%%%%%%%%%%%%%%%%%%%%%%%%%%%%%%%%%%%%%%%%%%%%%%%%%%%%%%%%

\section*{Presentation of the Institut Laue-Langevin (ILL)}

\begin{figure}[h!]
\begin{center}
\includegraphics[width=80mm]{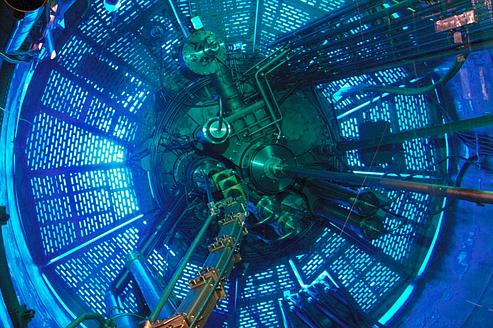}
\end{center}
\caption[Top of the 58.3 MW ILL core.] {Top of the 58.3 MW ILL core. Blue light emanating from it is caused by the Cherenkov effect.}
\label{IllCore}
\end{figure}

The Institut Laue-Langevin (ILL) is a research centre founded in 1967 and located in Grenoble, France. This facility, known worldwide for being the brightest continuous source of neutrons, is operated by three countries: Germany, France and the United Kingdom. Every year, the ILL welcomes approximately 1500 researchers and more than 800 experiments. This nuclear reactor (Fig. \ref{IllCore}) provides very high flux of neutrons to about 40 instruments that can be used for many different fields of research, from biology to particle physics for example. One of them is especially dedicated to particle and nuclear physics experiments: PF1b. It provides the strongest cold neutron beam in the world for such experiments.

My work placement has taken place in the Nuclear and Particle Physics group, in the Science division of the ILL and under the supervision of Dr Torsten \textsc{Soldner}. It consisted in installing the $a$SPECT experiment in the PF1B experimental zone, testing and creating software for the new acquisition system and participating in a lot of different analysis (related or not to the new acquisition system).

%%%%%%%%%%%%%%%%%%%%%%%%%%%%%%%%%%%%%%%%%%%%%%%%%%%%%%%%%%%%%%%%%%%%%%%%%%%%%%%
%  Théorie
%%%%%%%%%%%%%%%%%%%%%%%%%%%%%%%%%%%%%%%%%%%%%%%%%%%%%%%%%%%%%%%%%%%%%%%%%%%%%%%

\section{Theory}
\label{theorie}

Neutrons are electrically neutral baryons composed of two quarks $up$ and one quark $down$ and may possess a non-zero electric dipole moment (EDM)\footnote{In the SM the predicted neutron EDM is very small, $|d_n| \sim 10^{-32}e.cm$ \cite{Dar00}. For beyond the SM physics like SUper SYmmetry, the prediction is between $10^{-25}e.cm$ and $10^{-28}e.cm$ \cite{AKL01, PR05}. The actual upper limit is $|d_n| < 2.9\times10^{-26}e.cm$ \cite{Bak06}. Thus nEDM experiments have a privileged access to beyond the SM physics.} is possible. They are stable	when part of a nucleus, but a free neutron will decay into a proton, an electron and an electron-antineutrino with a life time $\tau_{n}$ of $880.0(9) s$ \cite{PDG12}:
\begin{eqnarray}
n \stackrel{\tau_{n}}{\rightarrow} p + e^{-} + \bar{\nu_{e}} + 782.3 \, {\rm keV}
\end{eqnarray}

with the released energy being the difference of mass between the neutron and the decay products. This decay is noted $\beta^{-}$ : the minus sign corresponds to the charge of the emitted charged lepton which is in this case an electron. The study of neutron beta decay is a very active field as it can lead, granted high-precision measurements, to beyond the SM physics (such as SUper SYmmetry or testing the conserved vector current theory (CVC)\footnote{This theory, proposed by Richard \textsc{Feynman} and Murray \textsc{Gell-Mann}, is based on an analogy with electromagnetism. Since the observed coupling strength $e$ with the electromagnetic field is the same for all coupled particles, the electromagnetic current is conserved. Then, if the weak vector current is similarly conserved, the vector coupling constant would be a universal constant \cite{Wu64}.} for example).

\begin{figure}
\begin{center}
\includegraphics[width=140mm]{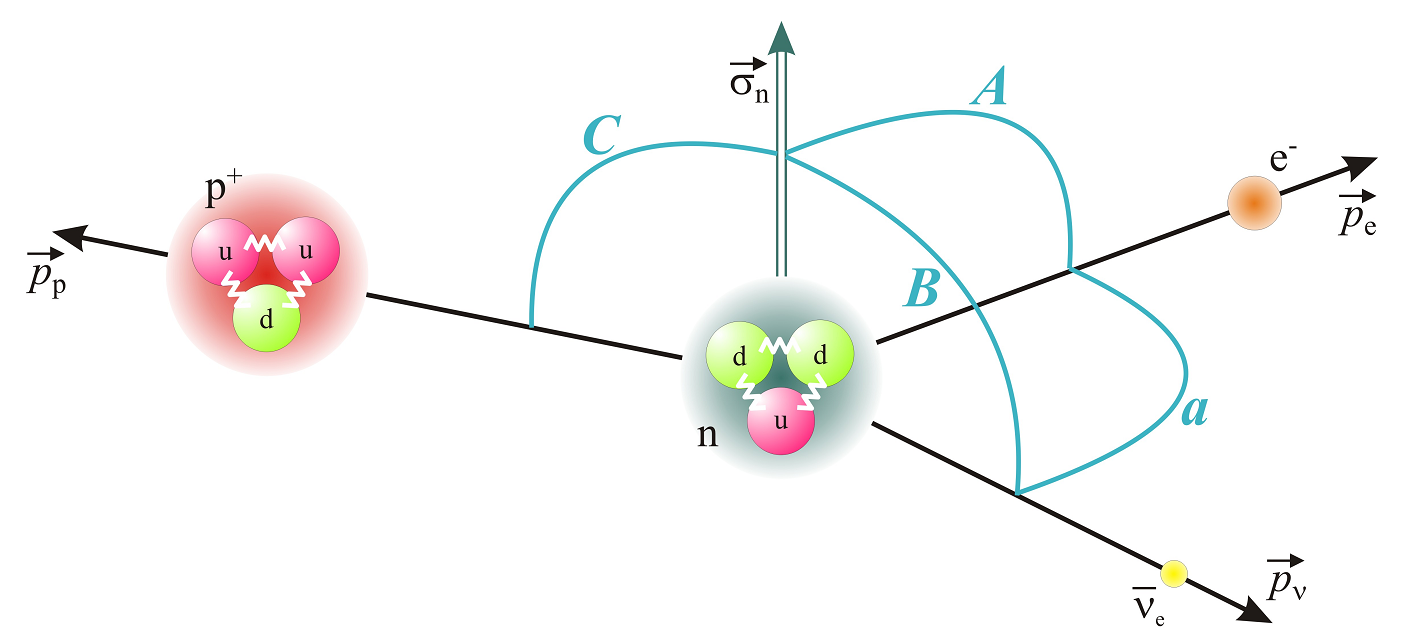}
\end{center}
\caption[Representation of the different angular correlation coefficients.]{Representation of the different angular correlation coefficients $a$, $A$, $B$ and $C$, where $\vec{\sigma_{n}}$ is the neutron spin. Figure taken from \cite{Kon11}.}
\label{angularcoef}
\end{figure}

The triple differential decay rate of neutrons can be written as \cite{JTW57}:

\begin{multline}
d^{3}\Gamma = \frac{1}{(2\pi)^{5}}\frac{G_{F}^{2}|V_{ud}|^{2}}{2}p_{e}E_{e}\,(E_{0}-E_{e})^{2}\,dE_{e}d\Omega_{e}d\Omega_{\nu} \\ \times \xi\left[1+a\frac{\textbf{p}_{e}\,.\,\textbf{p}_{\nu}}{E_{e}E_{\nu}}+b\frac{m_{e}}{E_{e}}+\frac{\langle \textbf{s}_{n}\rangle}{s_{n}}\left(A\frac{\textbf{p}_{e}}{E_{e}}+B\frac{\textbf{p}_{\nu}}{E_{\nu}}+D\frac{\textbf{p}_{e} 	\times\textbf{p}_{\nu}}{E_{e}E_{\nu}}\right)\right]
\label{eqn}
\end{multline}

where $G_{F}$ is the Fermi weak coupling constant, $V_{ud}$ is the upper left element of the Cabibbo-Kobayashi-Maskawa (CKM) matrix\footnote{The CKM matrix, also known as the quark-mixing matrix, informs on the strength of flavour-changing weak decays of quarks.}, $\textbf{p}_{e}$, $\textbf{p}_{\nu}$, $E_{e}$ and $E_{\nu}$ are respectively the electron (neutrino) momenta and total energies, $E_{0}$ is the electron spectrum endpoint total energy, $m_{e}$ is the electron mass, $\textbf{s}_{n}$ is the neutron spin and $\Omega_{i}$ correspond to the solid angles. $\xi$ is a factor which is inversely proportional to the neutron decay rate. $a$, $A$, $B$ and $D$ are the angular correlation coefficients and $b$ is the Fierz interference term. In the SM framework, $b=0$\footnote{$b$ is dependent of scalar and tensor couplings. In the SM, free neutron $\beta$ decay is only subject to vector and axial-vector couplings, resulting in $b=0$.}.

We can also define the coefficient $C$, the proton asymmetry relative to the neutron spin, as:
\begin{eqnarray}
C = -x_{C}(A+B)
\end{eqnarray}

with $x_{C}=0.27484$ being a kinematical factor. The different angular correlation coefficients are represented in Fig. \ref{angularcoef} and are the only parameters from Eq. (\ref{eqn}) that we'll consider for the rest of this report. Within the framework of the SM, we can express the angular correlation coefficients $a$, $A$, $B$ and $C$ as :
\begin{eqnarray}
a = \frac{1-|\lambda|^{2}}{1+3|\lambda|^{2}}, \;\;\;\;\;\;\;\;\; A = -2\frac{|\lambda|^{2}+\lambda}{1+3|\lambda|^{2}}, \;\;\;\;\;\;\;\;\; B = 2\frac{|\lambda|^{2}-\lambda}{1+3|\lambda|^{2}}, \;\;\;\;\;\;\;\;\; C = x_{C}\frac{4\lambda}{1+3|\lambda|^{2}} .
\label{eq4}
\end{eqnarray}

where $\lambda=L_{A}/L_{V}$ is the ratio of the weak axial-vector ($L_{A}G_{F}V_{ud}$) to the vector coupling constant ($L_{V}G_{F}V_{ud}$). The weak coupling constants are quite important as their size matter in several physics fields, like astronomy, cosmology or particle physics. 

On a side note, Eq. (\ref{eq4}) permits to cross-check the results of different experiments. Furthermore, physics beyond the SM enters differently into $a$, $A$, $B$ and $C$, hence new parameters would appear in Eq. (\ref{eq4}). As a result, disagreements for the value of $\lambda$ from $a$, $A$, $B$ and $C$ could indicate new physics. 

The sensitivities of $a$, $A$, $B$ and $C$ toward $\lambda$ near the world average value ($\lambda=-1.2701(25)$ \cite{PDG12}) are \cite{Kon11}:
\begin{eqnarray}
\frac{{\rm d}a}{{\rm d}\lambda}=0.298, \;\;\;\;\;\;\;\;\; \frac{{\rm d}A}{{\rm d}\lambda}=0.374, \;\;\;\;\;\;\;\;\; \frac{{\rm d}B}{{\rm d}\lambda}=0.076, \;\;\;\;\;\;\;\;\; \frac{{\rm d}C}{{\rm d}\lambda}=-0.124 .
\end{eqnarray}

We can see that $A$ is the most sensitive to $\lambda$, making it in theory the best parameter when it comes to the determination of $\lambda$. The Perkeo II experiment measured the most precise value for $A=-0.11972(_{-65}^{+53})$ \cite{Mun13}. The actual Particle Data Group value does not yet take this value into account and the world average value is $A=-0.1176(11)$ \cite{PDG12}. Since polarized neutrons are required in order to measure $A$ and the average neutron beam polarization is difficult to measure with a precision better than $10^{-3}$, it's difficult to decrease the error bars on $A$. The coefficient $a$, even though a little less sensitive to $\lambda$, does not require polarized neutrons and allows a systematically independent measurement of $\lambda$. Still, the present value of $a=-0.103(4)$ \cite{PDG12} is known with a relative uncertainty of $5\%$ and no real improvements have been made on it since 40 years. Hence improving the uncertainty in $a$ represents a major interest.

\begin{figure}
        \centering
        \begin{subfigure}{0.5\textwidth}
                \centering
                \includegraphics[width=\textwidth]{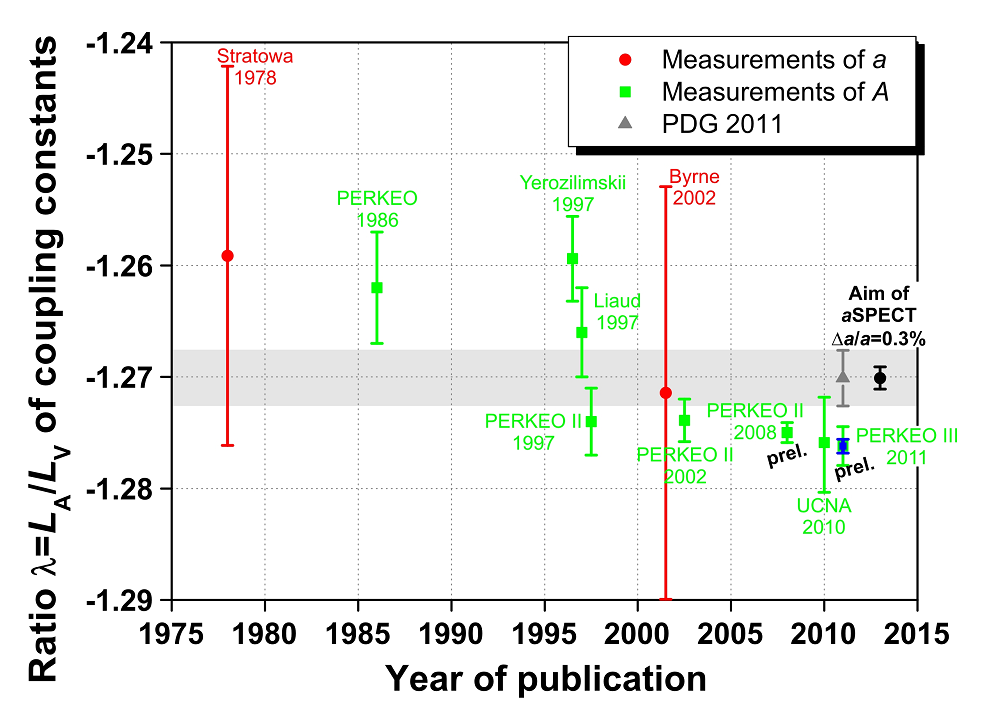}
                \caption{}
                \label{fig:aSPECTaim}
        \end{subfigure}%
        \begin{subfigure}{0.5\textwidth}
                \centering
                \includegraphics[width=\textwidth]{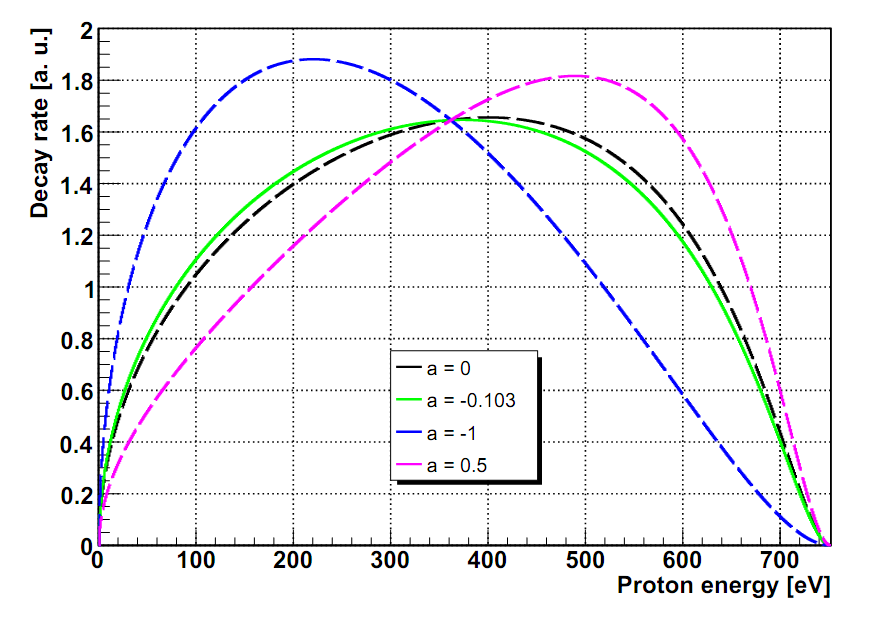}
                \caption{}
                \label{fig:protonspectrum}
        \end{subfigure}
				\vspace{0.5cm}
        \caption[(a): Values of $\lambda$ derived from measurements of $a$ and $A$ and (b): Predicted proton spectrum depending of the value of $a$.]{(a): Values of $\lambda$ derived from measurements of $a$ and $A$ over the last decades. The rightmost black circle represents what a $0.3\%$ measurement of $a$ would add to the determination of $\lambda$. Figure taken from \cite{Kon11}. (b): Predicted proton spectrum depending of the value of $a$. Figure taken from \cite{Sim10}.}\label{fig:aSPECT}
\end{figure}

The $a$SPECT experiment has been created in order to realise such high-precision measurement of $a$ with an announced aim of reducing the uncertainties from $5$ to $0.3\%$ (Fig. \ref{fig:aSPECTaim}). Since neutrinos are very hard to detect, the value of $a$ will be derived from the proton recoil spectrum, as shown in Fig. \ref{fig:protonspectrum}.

%%%%%%%%%%%%%%%%%%%%%%%%%%%%%%%%%%%%%%%%%%%%%%%%%%%%%%%%%%%%%%%%%%%%%%%%%%%%%%%
%  L'expérience aSPECT
%%%%%%%%%%%%%%%%%%%%%%%%%%%%%%%%%%%%%%%%%%%%%%%%%%%%%%%%%%%%%%%%%%%%%%%%%%%%%%%

\section{The \textit{a}SPECT experiment}
\label{sect:aSPECT}

% Présentation de l'expérience aSPECT

\begin{figure}
\begin{center}
\includegraphics[width=120mm]{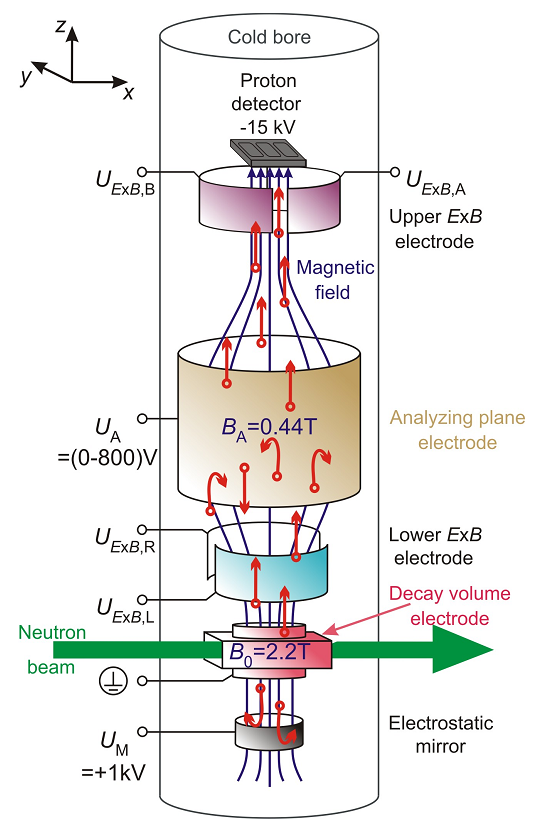}
\end{center}
\caption[Scheme of the $a$SPECT experiment.]{Scheme of the $a$SPECT experiment. A fraction of neutrons from the unpolarized, cold neutron beam decay in the decay volume (pink). Produced charged particles are then guided by the magnetic field either upward or downward. Protons (red) going downward are reflected by an electrostatic mirror (black). A first dipole electrode (turquoise) allows to sweep out trapped protons from the flux tube. A variable barrier voltage $U_{{\rm A}}$ in the analysing plane (gold) permits to scan the proton spectrum shape. The second dipole electrode (purple) serves to align the proton beam on the detector. Finally the high negative voltage post-accelerates protons. Figure taken from \cite{Kon11}.}
\label{aspectscheme}
\end{figure}

$a$SPECT is a retardation spectrometer, meaning that the spectrum is measured by counting the decay protons that overcome a defined potential barrier. This device (Fig. \ref{aspectscheme}) uses an unpolarized, cold neutron beam: about $10^{-8}$ of these free neutrons will decay in the decay volume (DV). Produced protons are then guided by a strong magnetic field (approximatively $2.2\,T$ in the DV) either towards the analysing plane (AP) (and thus the detector) or towards the opposite direction. In order to get the maximum statistic and to reduce systematic effects, protons emitted in the negative z-direction are reflected by an electrostatic mirror, resulting in a $4\pi$ detection solid angle. This electrostatic mirror is held at a positive voltage $U_{{\rm M}}>T_{{\rm p,max}}$ where $T_{{\rm p,max}}=751\,{\rm eV}$ \cite{SGB09} is the maximum kinetic energy of protons.
Protons gyrate around a magnetic field line and travel from the DV to the AP. 

In between those two elements, a dipole electrode creates an electric field perpendicular to the magnetic field. The Lorentz force acting on charged particles is expressed as:
\begin{eqnarray}
\textbf{F}=q(\textbf{E}+\textbf{v}\times \textbf{B})
\end{eqnarray}

resulting in a drifting velocity $\textbf{v}_{{\rm drift}}$:
\begin{eqnarray}
\textbf{v}_{\rm drift}=\frac{\textbf{E}\times \textbf{B}}{B^{2}} .
\end{eqnarray}

\begin{figure}
\begin{center}
\includegraphics[width=100mm]{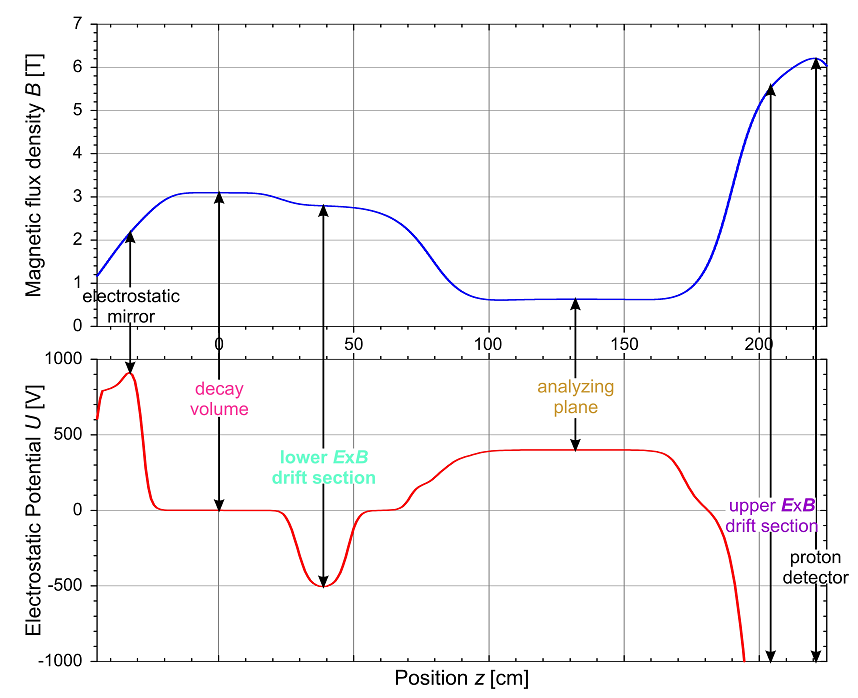}
\end{center}
\caption[Values of the magnetic field and electrostatic potential of the $a$SPECT experiment.]{Simulated values of the magnetic field (top) and electrostatic potential (bottom) along the z-axis of the $a$SPECT experiment. Note that this simulation has been realized with a higher current for the supraconducting coils. Actual magnetic field values are therefore roughly $30\,\%$ lower. Furthermore the lower $E\times B$ mean potential is $-100\,V$ for the 2013 beamtime.Figure taken from \cite{Kon11}.}
\label{fields}
\end{figure}

This drift, called E$\times$B (E-cross-B), is perpendicular to both magnetic and electric field. It is used to sweep out (after many passages) decay protons that would otherwise stay trapped between the AP and the DV (Fig. \ref{fields}). 

In the AP, a variable barrier voltage $U_{{\rm A}}$ is applied and only the protons with sufficient kinetic energy can pass this potential. $U_{{\rm A}}$ can vary from $0$ to $+800\,V$ in order to scan the proton spectrum shape. We can define the ratio of the magnetic fields in the AP and the DV as $r_{{\rm B}}=B_{{\rm A}}/B_{0}$ with $r_{{\rm B}}\approx1/5$ \cite{SGB09}. As a result of the adiabatic invariance of the magnetic moment\footnote{If the relative spatial changes of the magnetic and electric field stay small during one particle gyration ($\Delta B/B << 1$ and $\Delta E/E << 1$), a transfer of transverse to the field momentum into parallel momentum will conserve the total kinetic energy of this particle. The $a$SPECT design implies that all decay protons fulfill the condition for adiabatic transport. For example, the AP is preceded by several electrodes at lower potential to assure a smooth increase from 0 to $U_{{\rm A}}$, see Fig. \ref{fields}}, part of the proton\'s momentum transverse to the field is transferred into parallel momentum. 

We can define in the adiabatic approximation the transmission function $F_{{\rm tr}}(T)$ that represents the probability for a proton to overcome the potential barrier as \cite{SGB09}:
\begin{eqnarray}
F_{{\rm tr}}(T) = \left\{
\begin{array}{lr}
0 \;\;\;\;\;\;\;\;\;\;\;\;\;\;\;\;\;\;\;\;\;\;\;\;\;\;\;\;\;\;\;\;\;\;\;\;\;\;\;\, ; T<eU_{A}\\
1-\sqrt{1-\left(1-\frac{eU_{A}}{T}\right)/r_{B}} \;\;\;  ; {\rm otherwise}\\
1 \;\;\;\;\;\;\;\;\;\;\;\;\;\;\;\;\;\;\;\;\;\;\;\;\;\;\;\;\;\;\;\;\;\;\;\;\;\;\;\, ; T>\frac{eU_{A}}{1-r_{B}}
\label{trfunct}
\end{array}
\right.,
\end{eqnarray}

with $T$ being the kinetic energy and $e$ the charge of the proton.
Protons that have gone through the analysing plane are accelerated by a high voltage of $-15$ kV and are focused by a strong magnetic field onto a silicon drift detector (SDD). The second dipole electrode (Upper ExB) in between the AP and the SDD serves to align the proton beam on the detector. The high negative potential of the detector is required for post-accelerating the protons to detectable energies and to ensure that the protons overcome the magnetic mirror created by the increasing field (see section below). 

\subsection{The magnetic mirror effect}

\begin{figure}[b]
\begin{center}
\includegraphics[width=100mm]{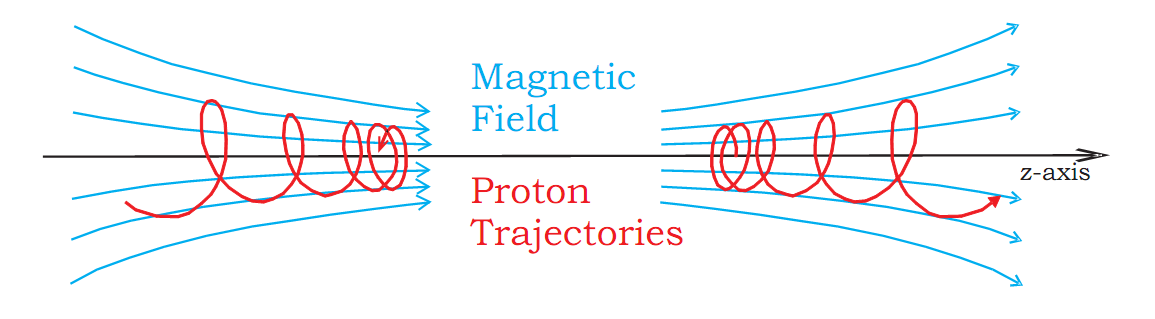}
\end{center}
\caption[Magnetic mirror effect representation.]{Magnetic field lines and proton trajectories in the normal (left) and the inverse (right) magnetic mirror effect. Figure taken from \cite{Sim10}.}
\label{magmirror}
\end{figure}

This phenomenon, as shown in Fig. \ref{magmirror}, can be divided into two cases :
\paragraph{Magnetic mirror effect:}
This effect corresponds to the fact that if a particle is moving towards an increasing magnetic field, its velocity component parallel to this field decreases. In order to focus to a small detector area ($\approx1$ cm$^{2}$), there is a strong magnetic mirror effect near the detector: a high negative potential is needed in order for the protons to overcome this effect.
\paragraph{Inverse magnetic mirror effect:}
On the other hand, if a particle is moving towards a decreasing magnetic field, its velocity component parallel to this field increases. As the potential barrier of the AP is only sensitive to the longitudinal energy component of the proton, an inverse magnetic mirror  allows to transfer practically all of the proton's momentum to the parallel component. Because of the adiabatic invariance of the magnetic moment, the total kinetic energy of the particle is not changed during this process. This effect is used on the AP, as shown in Fig. \ref{fields}.

\subsection{The silicon drift detector (SDD)}

A silicon drift detector is a semiconductor detector based on the principle of sidewards depletion, allowing the depletion of a large detector volume with a small readout node. As the thermal noise is proportional to the total anode capacitance, the SDD has a greatly reduced noise compared to a conventional PIN diode, whose capacitance increases linearly with the active area. Furthermore, the integration of the first amplification FET\footnote{A Field-Effect Transistor (FET) is a transistor that controls the conductivity of a channel in a semiconductor material by using an electric field. It is commonly used as an amplifier.} directly onto the detector bulk minimizes stray capacitances and removes several issues related to wire connections. Finally because of the better proton signal/noise separation the SDD allows to significantly reduce the acceleration voltage in the $a$SPECT spectrometer compared to the previous used silicon PIN diode detector, avoiding problems like electrical breakdowns or instabilities of the background due to field emission.

\begin{figure}[h!]
\begin{center}
\includegraphics[width=90mm]{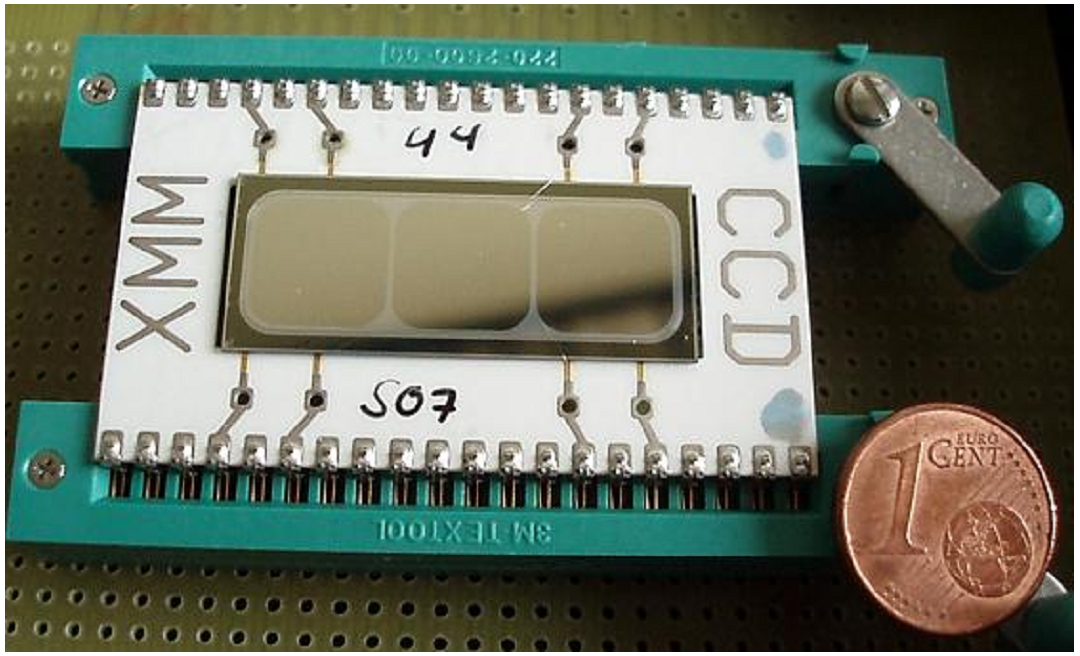}
\end{center}
\caption{The silicon drift detector chip of the $a$SPECT experiment.}
\label{chip}
\end{figure}

\begin{figure}
        \centering
        \begin{subfigure}{0.57\textwidth}
                \centering
                \includegraphics[width=\textwidth]{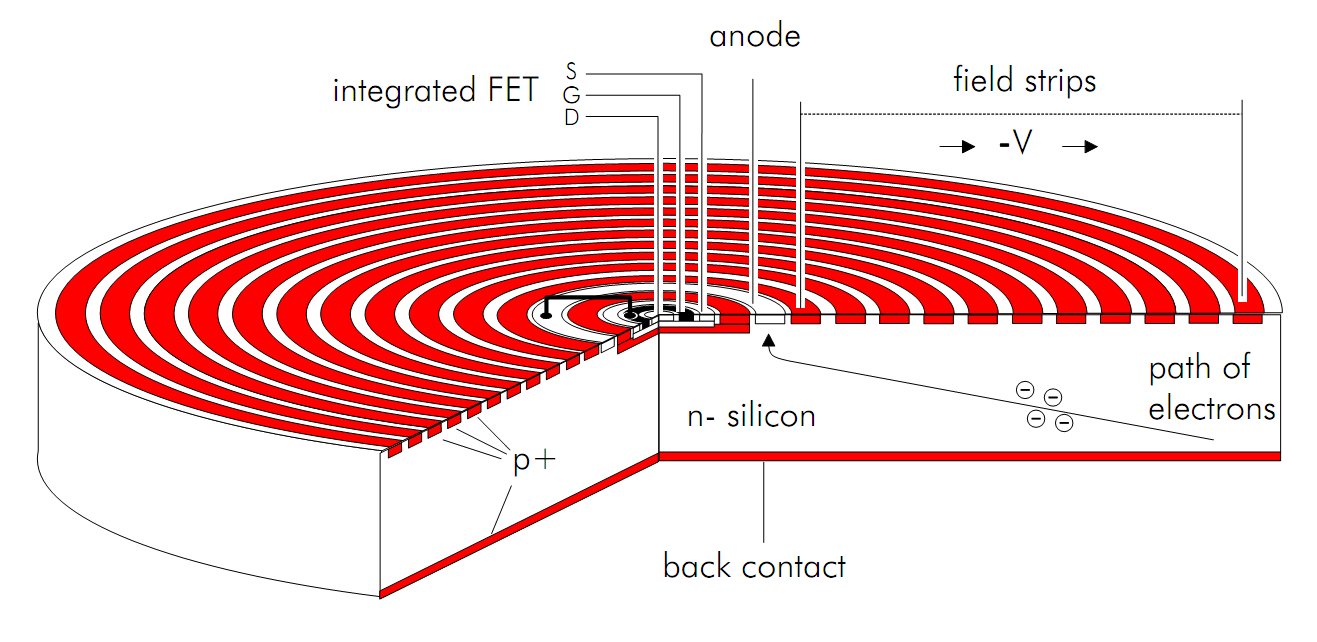}
                \caption{}
                \label{fig:sdd1}
        \end{subfigure}%
        \begin{subfigure}{0.35\textwidth}
                \centering
                \includegraphics[width=\textwidth]{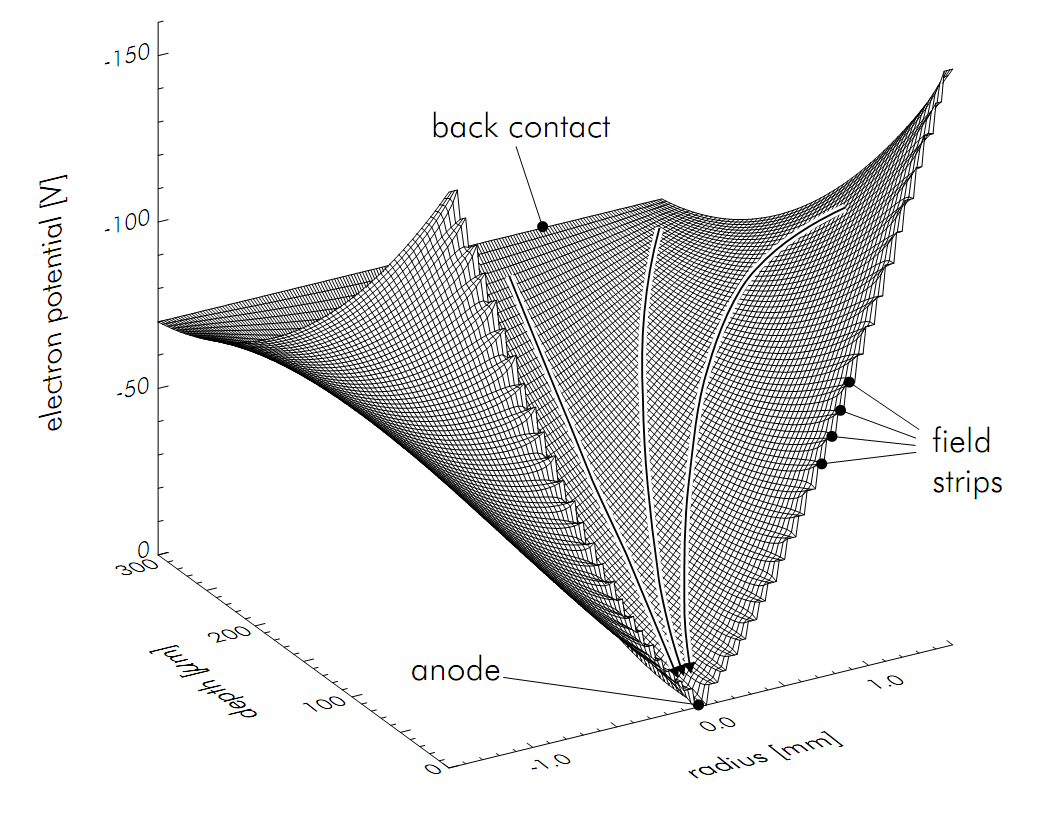}
                \caption{}
                \label{fig:sdd2}
        \end{subfigure}
				\vspace{0.5cm}
        \caption[(a): Schematic cross section of a SDD and (b): Calculated distribution of the potential energy in a SDD.]{(a): Schematic cross section of a SDD : radiation enters the detector through the back contact. Electrons are created by ionizing particles and drift towards a central ring-shaped anode. This anode is directly connected to an integrated first amplification stage (FET). (b): Calculated distribution of the potential energy in a SDD. This simulation applies to the cross section of a detector as shown in Fig. \ref{fig:sdd1} and the arrow lines indicate the signal electrons' drift directions \cite{LFL04}.}\label{fig:sdd}
\end{figure}

The chip used in $a$SPECT (Fig. \ref{chip}) is composed of 3 detector pads of $10 \times 10 \, {\rm mm}^{2}$ each. Protons enters the detector through the back contact (Fig. \ref{fig:sdd1}) and create, by ionisation, pairs of holes and electrons in the depletion layer. Holes travel to the cathodes (red parts in Fig. \ref{fig:sdd1}) and electrons drift along the potential valley (Fig. \ref{fig:sdd2}) towards the anode where they are collected. This potential valley is created by the cathodes : 73 rings with a voltage step of about 3 to 3.5 V from one ring to the next. Output signal of the detector is a negative pulse with a fast fall time followed by a very long exponential decay of the signal amplitude.

\subsection{Signal processing electronics}
\label{OldDAQSignalProcElec}

\begin{figure}
\begin{center}
\includegraphics[width=110mm]{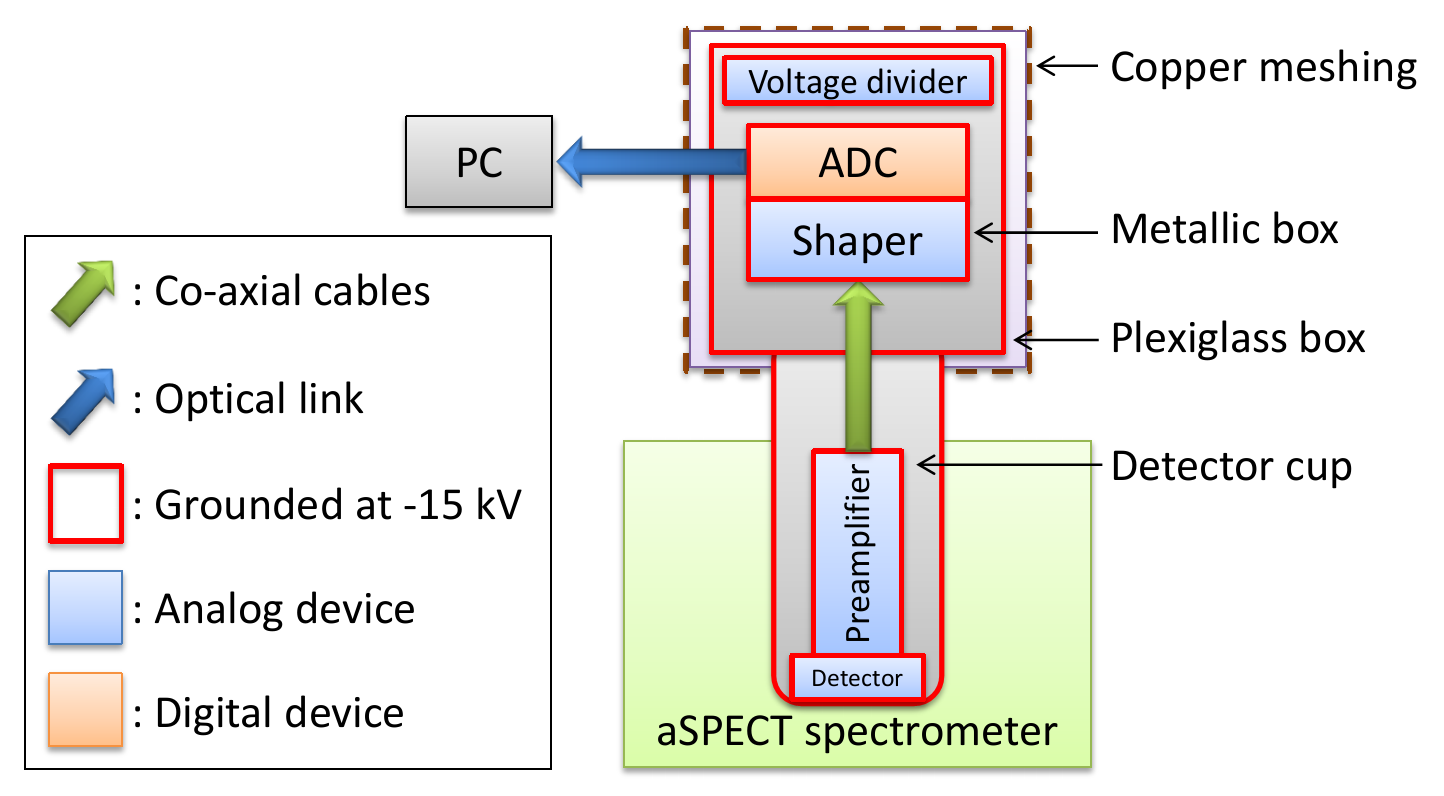}
\end{center}
\caption[Scheme of the electronics of $a$SPECT.] {Scheme of the $a$SPECT electronics. The only co-axial cables link the preamplifier output to the shaper input. The ADC is directly connected to the later and then an optical link is used to transfer triggered event data to a computer.}
\label{OldDAQScheme}
\end{figure}

Since the ground of the SDD is connected to the $-15$ kV potential, so is the ground of all the electronics handling the signal processing. The electrode where this high-voltage is applied is called the detector cup, as it is also the support for the SDD. The electrical isolation is done using an optical link between the final stage of the processing and the computer recolting the data. This section describes the acquisition system that has been used since the beginning of $a$SPECT (Fig. \ref{OldDAQScheme}). The new acquisition system will be detailed and studied in section \ref{newDAQ} $\,$ In order to distinguish the two different system easily, they will be respectively labelled as "old DAQ" and "new DAQ".

The electric alimentation of the detector and the different signal processing cards is handled by a voltage divider card, located in the electronics box at the top of $a$SPECT. This metallic box is grounded to the $-15$ kV and is inside a plexiglass box covered by a copper meshing. This lattice acts as a Faraday cage, blocking electromagnetic interferences.

\subsubsection{The preamplifier}

The preamplifier is the first board of this chain and is connected directly behind the detector in order to reduce the noise and loss of signal amplitude due to cables. This card will invert the signal and amplifies it through two levels of amplification (Fig. \ref{fig:PreampEvent}). The first one is separate for each of the three pads and is structured around Amptek A250 preamplifiers\footnote{Amptek A250 are state-of-the-art charge sensitive preamplifier with ultra low noise and high reliability screening. Amplification level can be adjusted by soldering or not pins corresponding to different capacitance values.}. The second one uses a single chip for the three channels: it is a 4 channels, high-speed and high output voltage amplifier.

Before July 2011, high-energy electron signals could easily saturate the preamplifier, leading to a miscalculation of the energy of the correlated proton (a detailed analysis of this problem can be found in \cite{Sim10}). In order to avoid the saturation, the amplification of the first stage was reduced.

\subsubsection{The shaper}

\begin{figure}[h!]
        \centering
        \begin{subfigure}{0.45\textwidth}
                \centering
                \includegraphics[width=\textwidth]{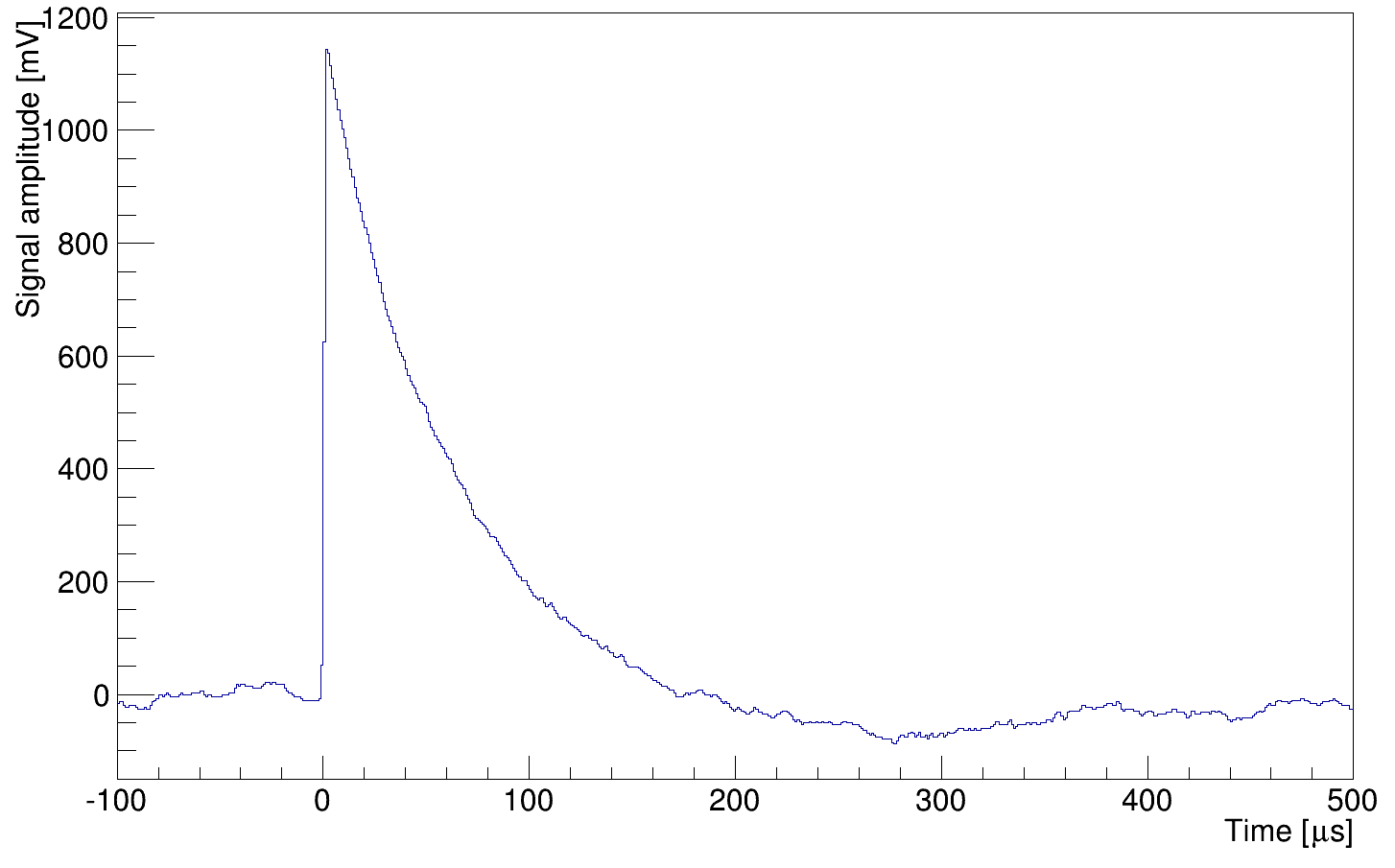}
                \caption{}
                \label{fig:PreampEvent}
        \end{subfigure}
        \begin{subfigure}{0.51\textwidth}
                \centering
                \includegraphics[width=\textwidth]{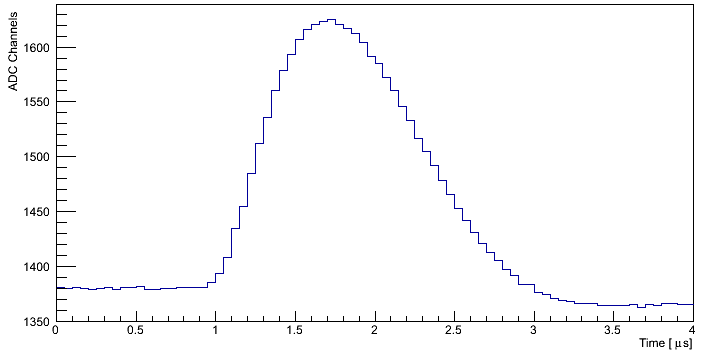}
                \caption{}
                \label{fig:NewShaperEvent}
        \end{subfigure}
				\vspace{0.5cm}
        \caption[(a): Cosmic event pulse after the preamplifier and (b):  Shaped event pulse after the shaper.]{(a): Pulse of a cosmic event after the preamplifier. Rise time of the signal is very fast compared to the very long exponential decay. (b): Event pulse after the shaper. The pulse length has been reduced from $150\, {\rm \mu s}$ to $2\, {\rm \mu s}$.}\label{fig:ShapersEvent}
\end{figure}

This analog component is located in the electronics box at the top of $a$SPECT (Fig. \ref{OldDAQScheme}). It is dedicacted to the shaping of the signal, transforming a very long pulse into a smaller one so it can be handled by the following Analog to Digital Converter (ADC). Since it is mostly sensitive to the rising part of the input pulse, the output pulse of the shaper will only last a few microseconds. As a result a $4\,{\rm \mu s}$ acquisition window is enough to get a single event pulse (Fig. \ref{fig:NewShaperEvent}).

For a linear preamplifier and shaper, the deposited energy of an event is directly proportionnal to the pulseheight. It is calculated by subtracting the baseline value to the maximum of the pulse. The baseline value is defined by the mean value of the 15 first bins of an event (first $0.75 \, \mu s$ of an event window).

\begin{figure}[h!]
        \centering
        \begin{subfigure}{0.48\textwidth}
                \centering
                \includegraphics[width=\textwidth]{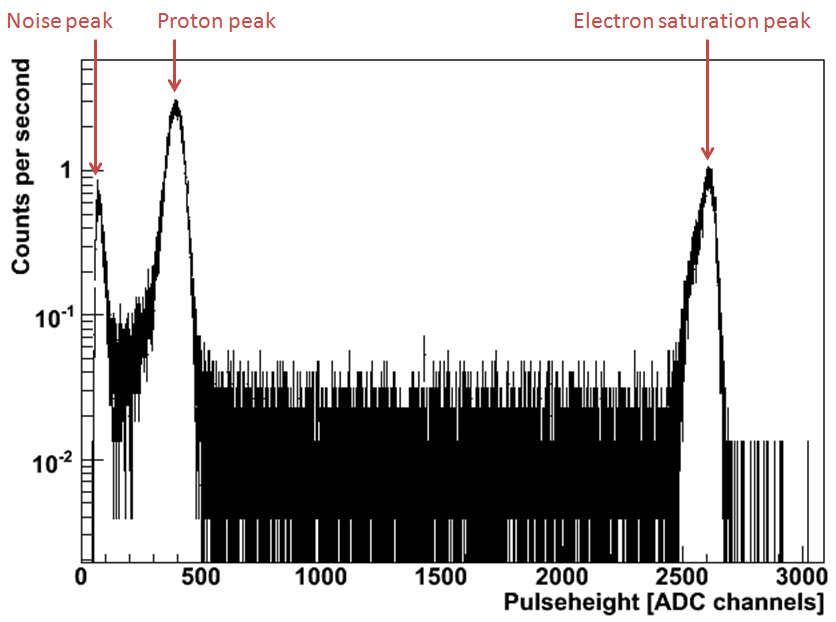}
                \caption{}
                \label{fig:OldShaper}
        \end{subfigure}
        \begin{subfigure}{0.5\textwidth}
                \centering
                \includegraphics[width=\textwidth]{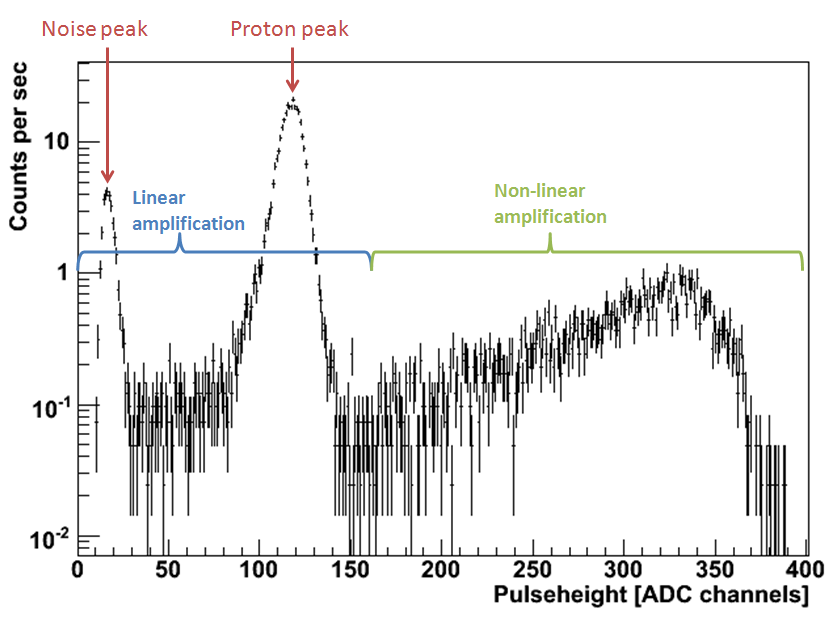}
                \caption{}
                \label{fig:NewShaper}
        \end{subfigure}
				\vspace{0.5cm}
        \caption[(a): Energy spectrum with old shaper design and (b):  Energy spectrum with new shaper design.]{(a): Energy spectrum with the old shaper design. Amplification is fully linear, resulting in a substantial saturation from high energy electrons. (b): With the new shaper design, the linear part is the same but the non-linear part "compresses" the electron spectrum. Both figures are taken from \cite{Mai13}.}\label{fig:Shapers}
\end{figure}

Like the preamplifier, the shaper had a saturation problem: high-energy electrons would cause the shaper to saturate (Fig. \ref{fig:OldShaper}). To solve this problem, a new design has been realized in July 2011. This new version has the particularity to apply a non-linear amplification above a certain signal amplitude (Fig. \ref{fig:NewShaper}). This limit should be above the proton event amplitude and thus only concern high-energy electrons. There are 6 output channels for this card, 2 per detector pad: one with high but non-linear amplification (like Fig. \ref{fig:NewShaper}) and one with linear but very small amplification.

\subsubsection{The Analog to Digital Converter (ADC)}

\begin{figure}
\begin{center}
\includegraphics[width=80mm]{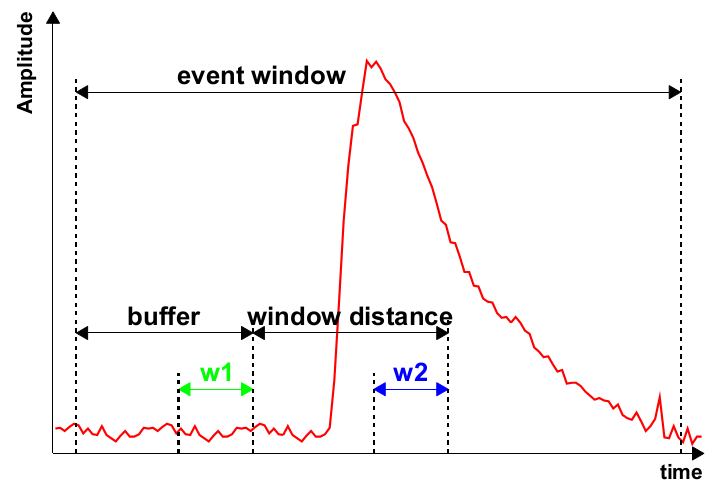}
\end{center}
\caption[Trigger algorithm of the old DAQ.] {Trigger algorithm scheme of the old DAQ.}
\label{OldDAQTrigger}
\end{figure}

The last part of the signal processing electronics is the ADC. Output signal of the 6 shaper channels are continuously digitalized with a $12$ bit resolution and a $20\,{\rm MHz}$ sampling frequency, resulting in a $50\,{\rm ns}$ time resolution. To reduce the amount of data to handle afterwards, two FPGA\footnote{A Field-Programmable Gate Array is an integrated circuit that can be configured by a customer after manufacturing, it is programmable.} treat the data using a trigger algorithm. This algorithm continuously compare two windows of the signal separated by a certain distance (Fig. \ref{OldDAQTrigger}). If the average value of the last window (w2) is higher than the mean value of the first one (w1) plus a treshold, than a trigger decision is made and the event is stored.

\clearpage

%%%%%%%%%%%%%%%%%%%%%%%%%%%%%%%%%%%%%%%%%%%%%%%%%%%%%%%%%%%%%%%%%%%%%%%%%%%%%%%
%  Experimental setup
%%%%%%%%%%%%%%%%%%%%%%%%%%%%%%%%%%%%%%%%%%%%%%%%%%%%%%%%%%%%%%%%%%%%%%%%%%%%%%%

\section{Experimental setup}
\label{Positions}

\begin{figure}[h!]
\begin{center}
\includegraphics[width=160mm]{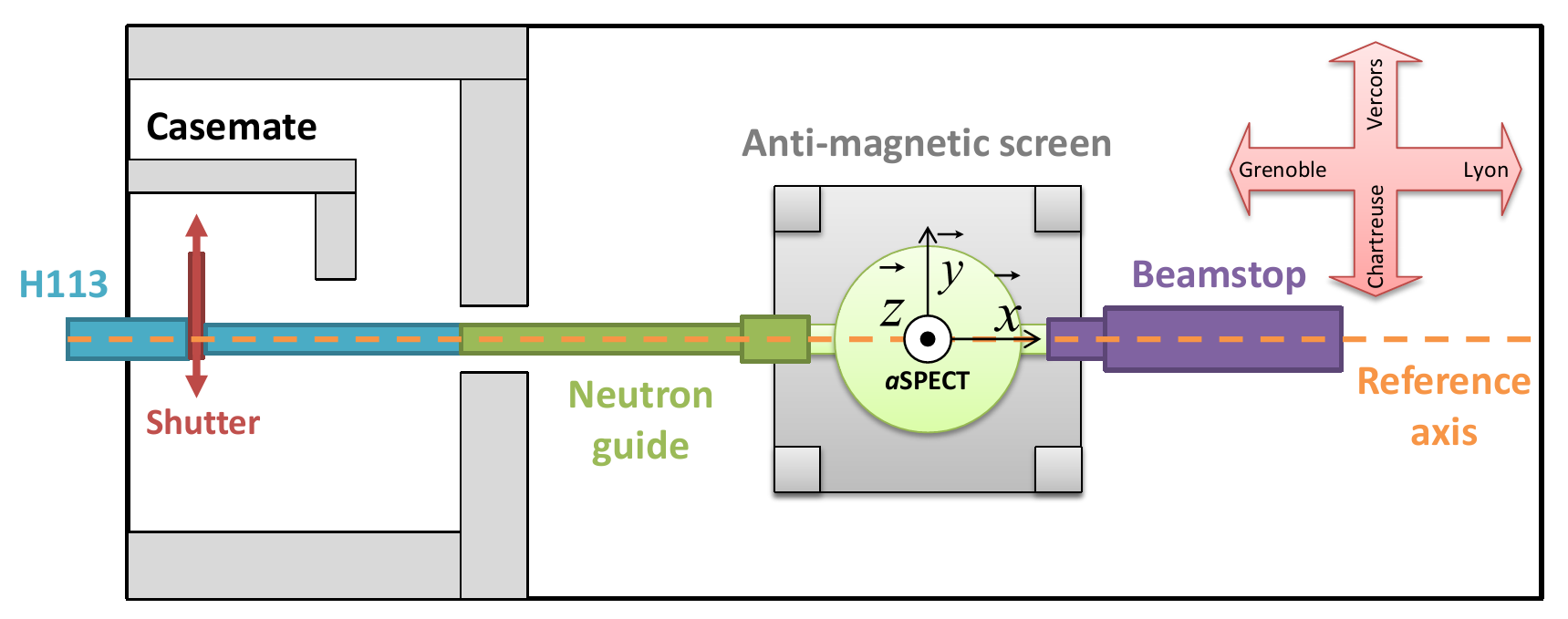}
\end{center}
\caption[Frame of reference for $a$SPECT and PF1B zone.] {Frame of reference for $a$SPECT and PF1b zone setup. Point $(0,0,0)$ is the center of the decay volume. The x-axis is parallel to the neutron beam and in the same direction as neutrons going out from the reactor. $\textbf{z}$ is the vertical axis: positive z pointing toward the sky and negative z toward the earth. It is a right-handed frame, thus positive y are pointing toward "Vercors". The reference axis is defined by the center of the H113 neutron guide.}
\label{aSPECTPositions}
\end{figure}

The detector position inside the $a$SPECT spectrometer and its projected area in the Decay Volume has to be known in order to run valid simulations of the experiment. Furthermore the neutron beam profile needs to be known precisely in order to determine the edge effect systematic uncertainty. 
The common frame of reference used for $a$SPECT and for this report is described in Fig. \ref{aSPECTPositions}. The center of this frame is the center of the decay volume.

\begin{figure}[h!]
\begin{center}
\includegraphics[width=80mm]{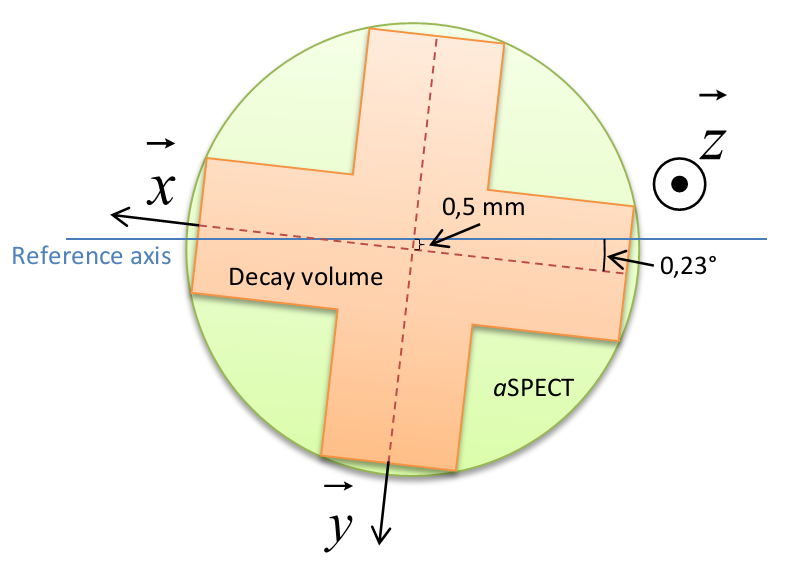}
\end{center}
\caption[Spectrometer alignement.] {Alignement of the $a$SPECT spectrometer compared to the reference axis. Orientation and displacement are not at scale.}
\label{aSPECTrotation}
\end{figure}

For the 2013 beamtime, the $a$SPECT experiment was installed in the PF1b experimental zone of the ILL. To align the spectrometer with the neutron guide we used a theodolite. This tool gives us the y-displacement of a position compared to the reference axis which is defined by the center of the neutron guide. Because of the important weight of the experiment it is not easy to align with this reference axis. The measured position of the center of the entrance and exit flange give us the position and alignement of the spectromter, as shown in Fig. \ref{aSPECTrotation}. 

\begin{figure}[h!]
\begin{center}
\includegraphics[width=100mm]{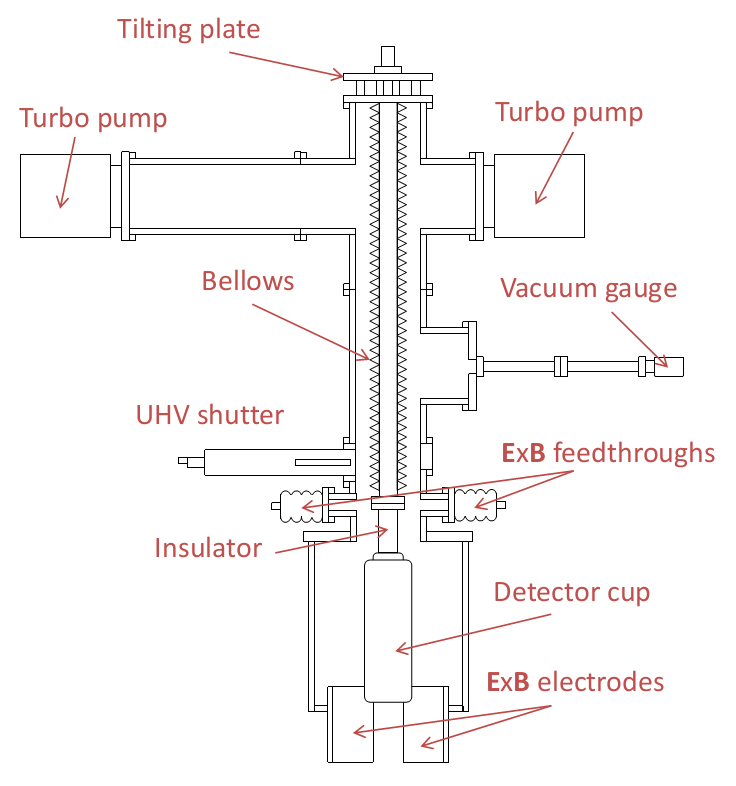}
\end{center}
\caption[Detector mechanics.] {Scheme of the detector mechanics and his retractable system. In this figure the detector is inserted inside the $a$SPECT spectrometer. When moved up, the detector cup is above the Ultra-High-Vacuum shutter, allowing to access the detector without venting the main vacuum.}
\label{MechanicalSetup}
\end{figure}

Because of the high magnetic field needed to focus the protons onto the SDD, the detector needs to be inside the $a$SPECT spectrometer. Since the detector and its electronics are sensitive elements of the experiment, an easy access to those parts is required if a maintenance is needed. Warming up and venting the cryostat is not a valid option as this process would require to pump and cool down again the spectrometer, which takes at least ten days. As a result a retractable system is used, allowing to access the detector without venting or warming up the main cryostat (Fig. \ref{MechanicalSetup}).

\begin{figure}
        \centering
        \begin{subfigure}{0.35\textwidth}
                \centering
                \includegraphics[width=\textwidth]{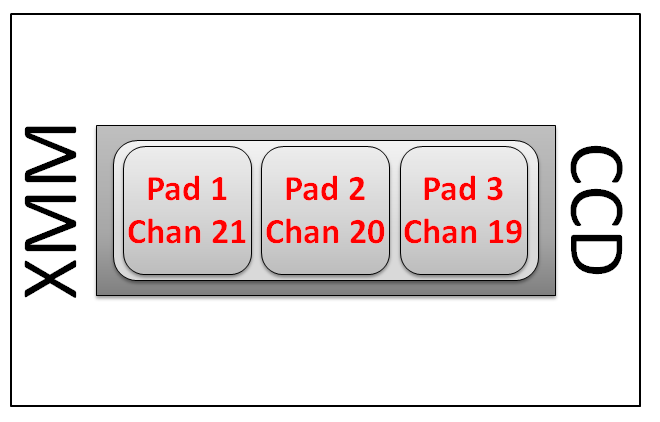}
                \caption{}
                \label{DetPads}
        \end{subfigure}
        \begin{subfigure}{0.6\textwidth}
                \centering
                \includegraphics[width=\textwidth]{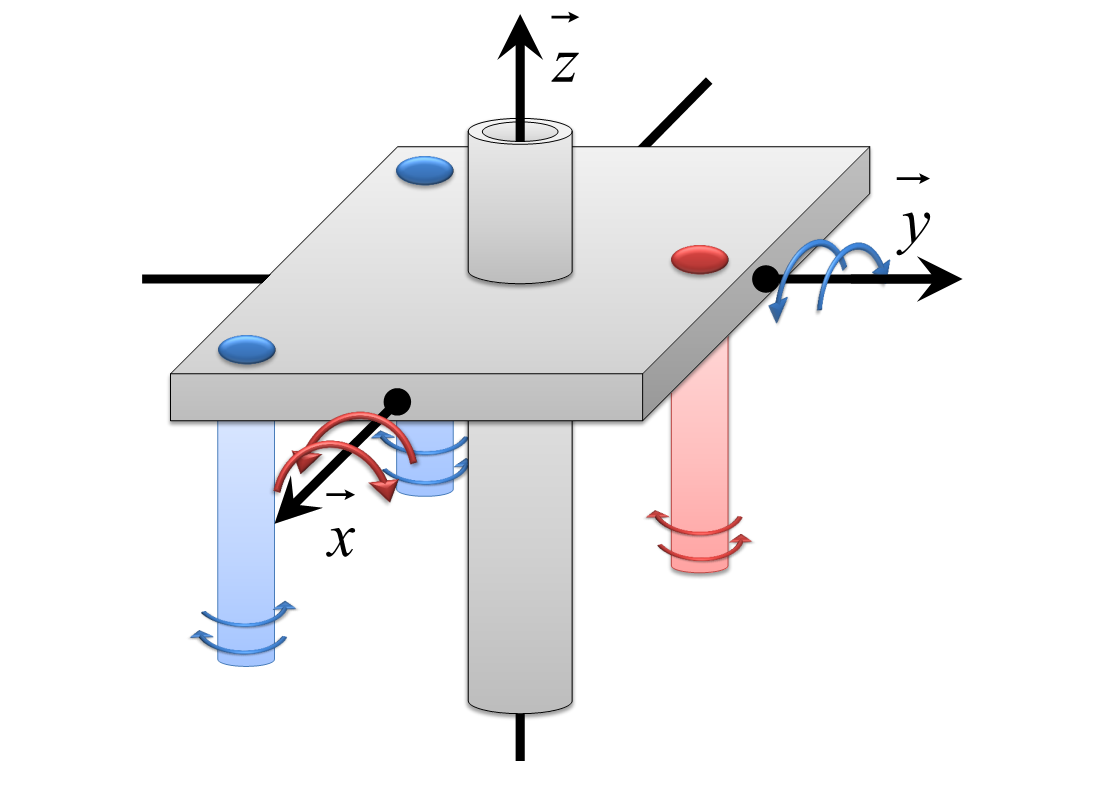}
                \caption{}
                \label{DetPosMech}
        \end{subfigure}
				\vspace{0.5cm}
        \caption[(a): Detector pads and channels and (b):  Tilting plate principle.]{(a): Position of the detector pads and corresponding channels on the acquisition system. (b): Scheme of the tilting plate. By turning the red screw, the plate is tilted around the x-axis, allowing to scan the y-direction. By turning both blue screws in opposite direction the plate is tilted around the y-axis, scanning the x-direction.}\label{fig:Shapers}
\end{figure}

Fig. \ref{DetPads} shows the position of the pads on the detector and their corresponding channels with the acquisition system. The inverted order of the channels comes from the shaper. The detector was oriented with pad 3 toward the reactor (Grenoble) and pad 1 toward the beam stop (Lyon).

\subsection{Detector position}
\label{DirectMeasurement}

In addition to the retractable system a new upper part has been developped by Dr. Martin \textsc{Simson}: it allows to move the detector in the xy-plane. As shown in Fig. \ref{DetPosMech}, three screws allow to incline the upper plate and thus to move the tube and the detector cup. However since this tube is quite long (approximately $130$ cm) and the detector cup quite heavy the reproducibility of the position is low. 

Several scans and tests have been realized with the detector mechanics outside and inside $a$SPECT. The result is that the reproducibilty of the detector position when moving up and down the detector is not perfect, and it is only reproducible within approximately $2.5\,{\rm mm}$. In order to get the most precise detector position, the position of the moved down detector cup has been measured with the detector mechanics outside $a$SPECT. The detector was then moved up, the detector mechanics installed on $a$SPECT and finally, after reaching sufficient vacuum conditions, the detector was moved down. Another uncertainty is the force applied by the ultra-high vacuum of the spectrometer. It is however not known nor really predicted.

The measured position with the detector moved down outside $a$SPECT is:

\begin{itemize}
	\item $-0.68(18)\,{\rm cm}$ shift in the x-direction: the detector cup is shifted toward Grenoble.
	\item $+0.18(18)\,{\rm cm}$ shift in the y-direction: the detector cup is slightly shifted toward Vercors.
\end{itemize}

Therefore this position is the one, within the reproducibilty uncertainty, of the center of the detector inside $a$SPECT.

\subsubsection{Detector orientation}
\label{DetectorOrientation}

\begin{figure}[h!]
\begin{center}
\includegraphics[width=130mm]{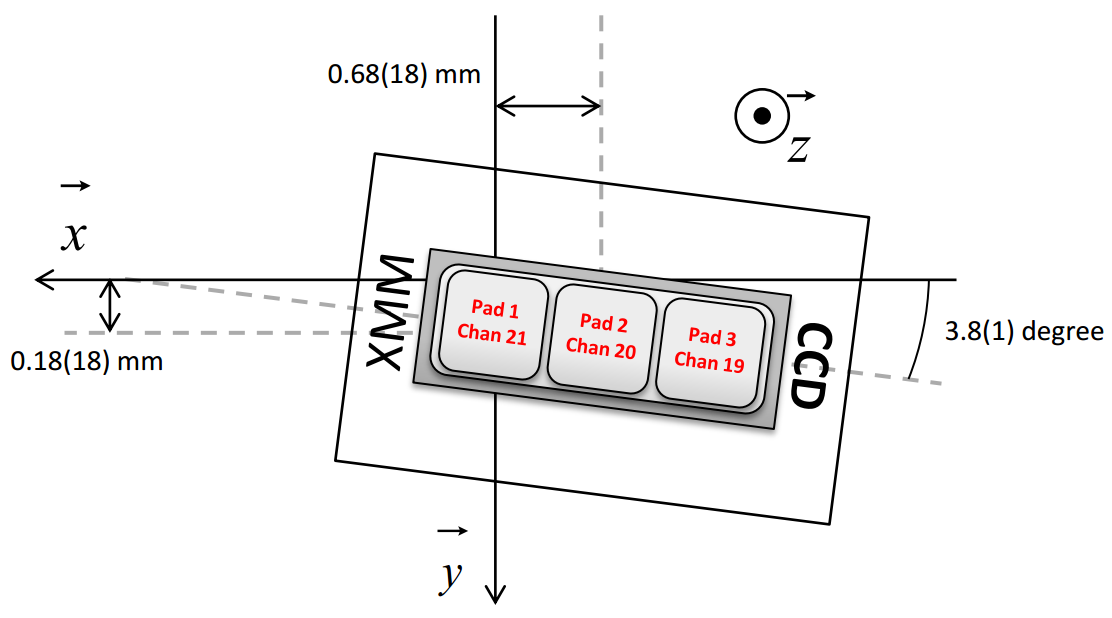}
\end{center}
\caption[Position of the detector inside the spectrometer.] {Position of the detector inside the spectrometer.}
\label{DetPosFinal}
\end{figure}

The SDD can also be tilted compared to the x-axis. A direct measurement of the detector orientation has been realized by marking the reference axis on the bottom flange of the detector mechanics. At the end of the beamtime the detector mechanics were removed and then a picture was taken. An offline analysis of this picture with Mathematica allowed to determine precisely the angle between the detector and the reference axis. The full detector position is described in Fig. \ref{DetPosFinal}.

\subsection{Detector projected area}
\label{CopperWireScan}

\begin{figure}[h!]
\begin{center}
\includegraphics[width=160mm]{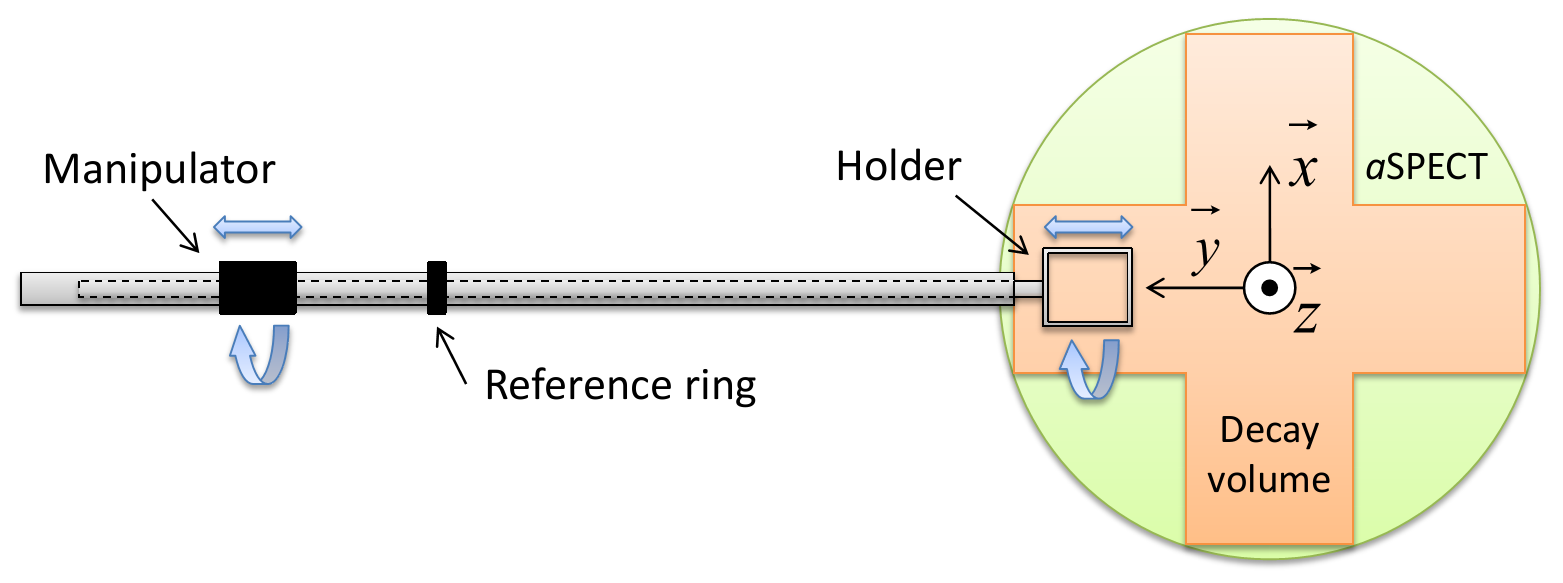}
\end{center}
\caption[Sample holder inside $a$SPECT.] {Scheme of the sample holder. By moving the manipulator along the axis or by turning it, one can move the sample fixed on the holder inside the decay volume. Graduations on both the translation and the rotation of the manipulator allows to choose the holder position and orientation.}
\label{CrossPieceHolder}
\end{figure}

In order to determine the detector projected area in the Decay Volume, we used a copper wire mounted on a holder fixed to a tube. Using a magnetic manipulator we can insert this tube (and thus samples) into the decay volume along the y-axis (Fig. \ref{CrossPieceHolder}).

\begin{figure}[h!]
\begin{center}
\includegraphics[width=80mm]{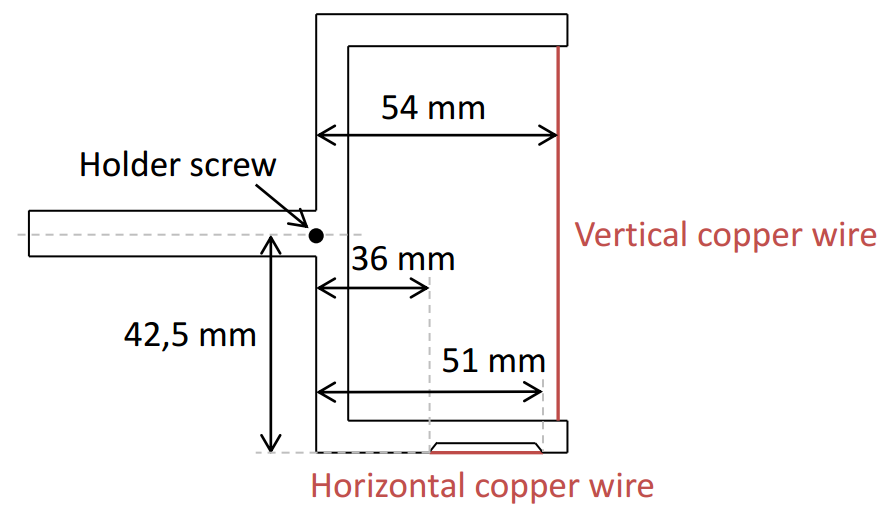}
\end{center}
\caption[Copper wire holder.] {Positions of the vertical and horizontal copper wire relative to the holder screw.}
\label{CuWireHolder}
\end{figure}

As shown in Fig. \ref{CuWireHolder}, a thin copper wire was fixed on a holder, parallel to the x-axis, and then activated using the neutron beam (more details on the copper activation in Sec. \ref{CopperFoilActivation}). By translating the manipulator, the y-axis can thus be scanned and the y-position of the detector projected area deduced from the resulting count rates. Furthermore, by placing a copper wire on the side of the holder parallel to the y-axis, one can scan the x-axis by turning the manipulator.

\paragraph*{X-scan}

\begin{figure}
        \centering
        \begin{subfigure}{0.49\textwidth}
                \centering
                \includegraphics[width=\textwidth]{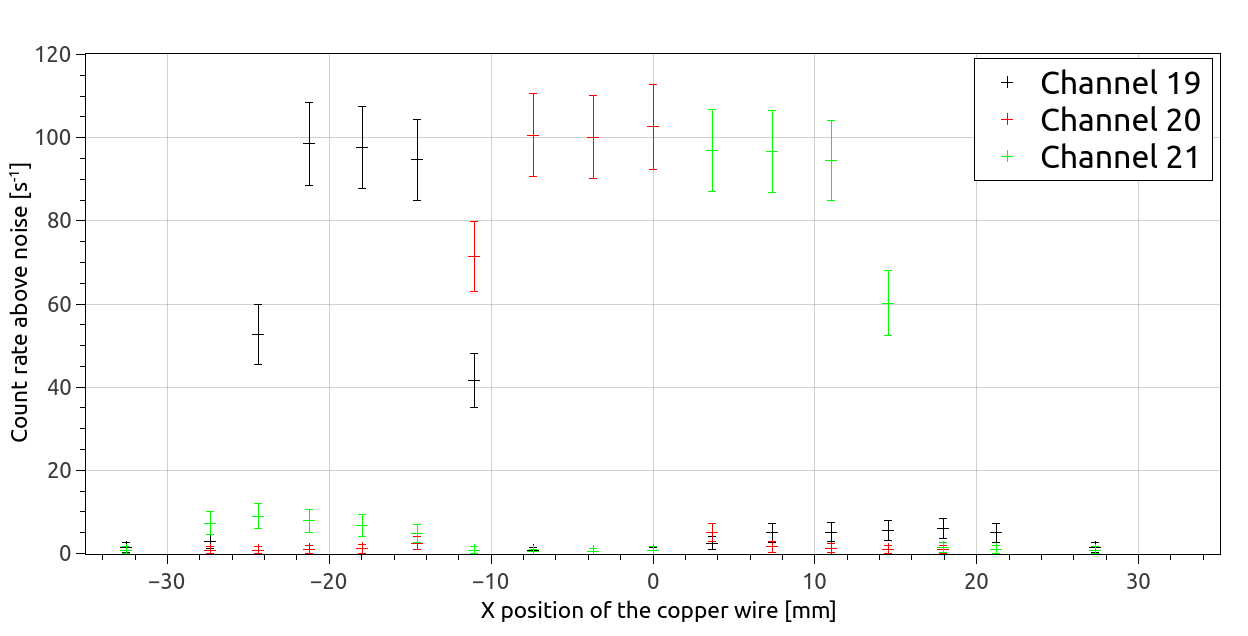}
                \caption{}
                \label{XscanCopperWire}
        \end{subfigure}
        \begin{subfigure}{0.49\textwidth}
                \centering
                \includegraphics[width=\textwidth]{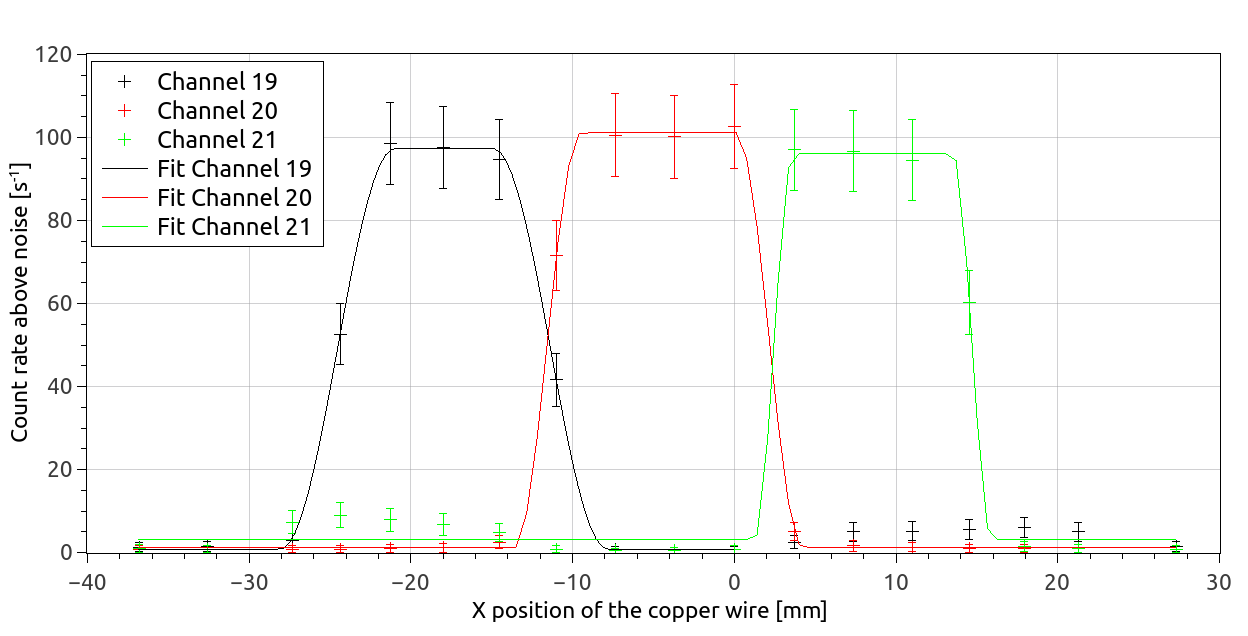}
                \caption{}
                \label{xCopperWireScanFitted}
        \end{subfigure}
				\vspace{0.5cm}
        \caption[(a): X-scan with the activated copper wire and (b): Fitting curves of the x-scan.]{(a): Count rate for each channel depending of the x-position of the activated copper wire. (b): Fit of the data using a phenomenological function.}\label{XWireScan}
\end{figure}

A manipulator angle of $-25\,{\rm deg}$ corresponds to the position $x=0$ for the horizontal copper wire. To convert the manipulator angle into the x-position, we use the following relation: $x=-\sin({\rm angle}+25\,{\rm deg})\times 42.5\,{\rm mm}$. Fig. \ref{XscanCopperWire} indicates the result of this scan.

A phenomenological function has been used to fit the data:

\begin{eqnarray}
F(x) = \left\{
\begin{array}{lr}
a_{0} \;\;\;\;\;\;\;\;\;\;\;\;\;\;\;\;\;\;\;\;\;\;\;\;\;\;\;\;\;\;\;\;\;\;\;\;\;\;\;\;\;\;\;\;\;\;\;\;\;\;\;\;\;\;\;\, ; x<{\rm LowerEdge}-{\rm EdgeWidth}\\
\vspace{0.2cm}
\frac{a_{0}+a_{1}}{2}+\frac{(a_{1}-a_{0})\times \sin(\frac{\pi}{2}\times \frac{x-{\rm LowerEdge}}{{\rm EdgeWidth}})}{2} \;\;\;\;\;\;\;\;\;\;\; ; x<{\rm LowerEdge}+{\rm EdgeWidth}\\
a_{1}\;\;\;\;\;\;\;\;\;\;\;\;\;\;\;\;\;\;\;\;\;\;\;\;\;\;\;\;\;\;\;\;\;\;\;\;\;\;\;\;\;\;\;\;\;\;\;\;\;\;\;\;\;\;\;\,  ; {\rm otherwise}\\
\vspace{0.2cm}
\frac{a_{0}+a_{1}}{2}-\frac{(a_{1}-a_{0})\times \sin(\frac{\pi}{2}\times \frac{x-{\rm HigherEdge}}{{\rm EdgeWidth}})}{2} \;\;\;\;\;\;\;\;\;\;\; ; x<{\rm HigherEdge}-{\rm EdgeWidth}\\
a_{0} \;\;\;\;\;\;\;\;\;\;\;\;\;\;\;\;\;\;\;\;\;\;\;\;\;\;\;\;\;\;\;\;\;\;\;\;\;\;\;\;\;\;\;\;\;\;\;\;\;\;\;\;\;\;\;\, ; x>{\rm HigherEdge}+{\rm EdgeWidth}
\label{xScanFct}
\end{array}
\right.,
\end{eqnarray}

Fig. \ref{xCopperWireScanFitted} shows the resulting fits. The poor angle resolution of this data set limits the precision of the analysis, therefore rough fit parameters had to be indicated for the fitting algorithm to converge.

As this fit function is truncated into several parts and the data points not numerous, uncertainties given for the fit parameters are huge and not reallistic. Therefore the position uncertainty is defined as $1.8\,{\rm mm}$ which is approximately half of the distance between each acquired point, as there is often a direct transition from 0 counts to the full count rate between two points. The scan results are summed-up in Table \ref{TabXscanWire}.

\begin{table}
	\centering
		\begin{tabular}{|c|c|c|}
			\hline
			\textbf{Pad} & \textbf{Detector projected} & \textbf{X-Width of the detector} \\
			 & \textbf{area center in x [mm]} & \textbf{projected area [mm]}\\
			\hline
			1 (Chan. 21) & $+8.6\pm 1.3$ & $12.27\pm 2.5$\\
			\hline
			2 (Chan. 20) & $-4.7\pm 1.3$ & $13.72\pm 2.5$\\
			\hline
			3 (Chan. 19) & $-18.0\pm 1.3$ & $13.17\pm 2.5$\\
			\hline
		\end{tabular}
		\vspace{0.5cm}
	\caption{Caracteristics of the detector projected area in the Decay Volume in the x direction, extracted from the activated copper wire scan.}
	\label{TabXscanWire}
\end{table}

%The measured shift of the detector position (Table \ref {TabXscanWire}) is of the same order of magnitude as the one measured in Sec. \ref{DirectMeasurement}. However the big uncertainty on both measurement (due on one side to the poor reproducibility of the detector insertion and on the other side to the lack of data points) does not give a clear result. A weighted average and error can be calculated as:

%\begin{eqnarray}
%\overline{x}\pm\delta\overline{x}=\frac{\sum_{i}\omega_{i}x_{i}}{\sum_{i}\omega_{i}}\pm \frac{1}{\sqrt{\sum_{i}\omega_{i}}}
%\end{eqnarray}

%with $\omega_{i}=1/ (\delta_{x_{i}})^{2}$.

%The weighted average value of the x-position of the center of the detector is: 

%\begin{center}
%$x_{center}=-5.4\pm1.1\,mm$
%\end{center}

\paragraph*{Y-scan}

The y-position of the projected area is determined by measuring the distance between the reference ring and the manipulator (see Fig. \ref{CrossPieceHolder}). The wire is aligned with the reference axis for a read value of $+33.5\,{\rm mm}$, meaning that the $y=0$ position for the copper wire corresponds to a read value of $34.0\,{\rm mm}$.

\begin{figure}[h!]
\begin{center}
\includegraphics[width=140mm]{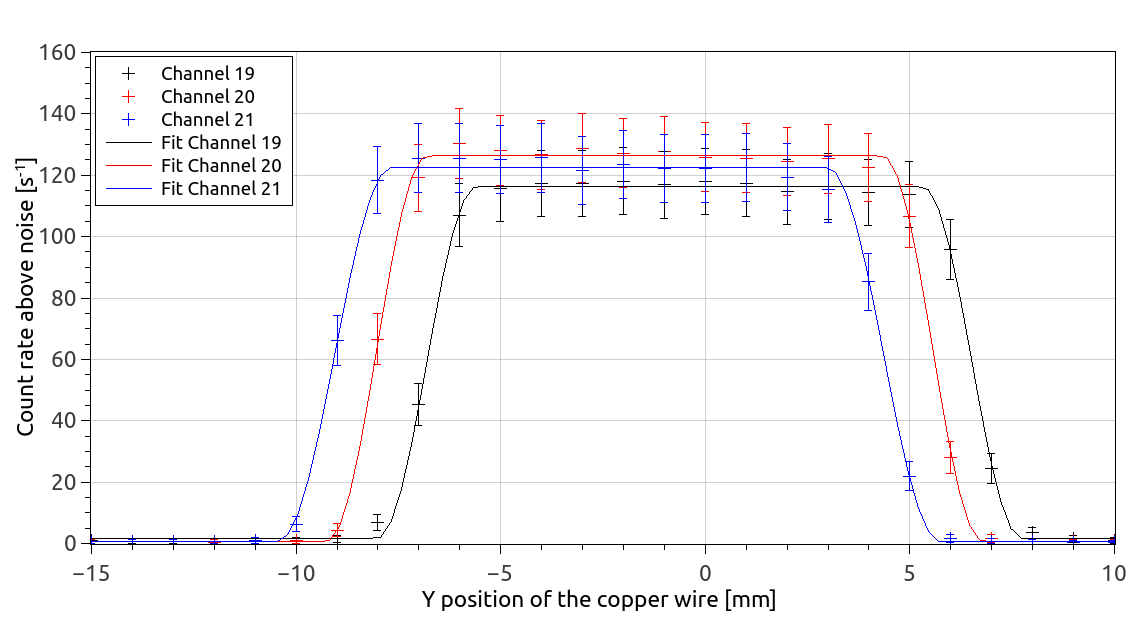}
\end{center}
\caption[Y-scan with the activated copper wire.] {Count rate for each channel depending of the y-position of the activated copper wire and their respective fit with the function (Eq. \ref{xScanFct}).}
\label{YscanCopperWire}
\end{figure}

The phenomenological fit function (Eq. \ref{xScanFct}) is also used for this data set (Fig. \ref{YscanCopperWire}): as there is a better spatial resolution compared to the x-scan, uncertainty of the fitting algorithm parameters are relevant.

The manipulator position is measured with a calliper rule with a $0.1\,{\rm mm}$ uncertainty and the edges positions given by the fit are known with a $0.02\,{\rm mm}$ uncertainty. As a result, by propagating the errors, the position of the center of the projected area is known with a $\sqrt{2\times(0.1^{2}+0.02^{2})}/2=0.07$ mm precision. Y-scan results are referenced in Table \ref{TabYscanWire}.

\begin{table}
	\centering
		\begin{tabular}{|c|c|c|}
			\hline
			\textbf{Pad} & \textbf{Detector projected} & \textbf{Y-Width of the detector} \\
			 & \textbf{area center in y [mm]} & \textbf{projected area [mm]}\\
			\hline
			1 (Chan. 21) & $-2.35\pm 0.07$ & $13.41\pm 2.5$\\
			\hline
			2 (Chan. 20) & $-1.23\pm 0.07$ & $13.58\pm 2.5$\\
			\hline
			3 (Chan. 19) & $-0.14\pm 0.07$ & $13.32\pm 2.5$\\
			\hline
		\end{tabular}
		\vspace{0.5cm}
	\caption{Caracteristics of the detector projected area in the Decay Volume in the y direction, extracted from the activated copper wire scan.}
	\label{TabYscanWire}
\end{table}

\subsection{Neutron beam profile}
\label{Neutronbeamprofile}

\subsubsection{Edge effect}
\label{EdgeEffectText}

\begin{figure}[h!]
\begin{center}
\includegraphics[width=160mm]{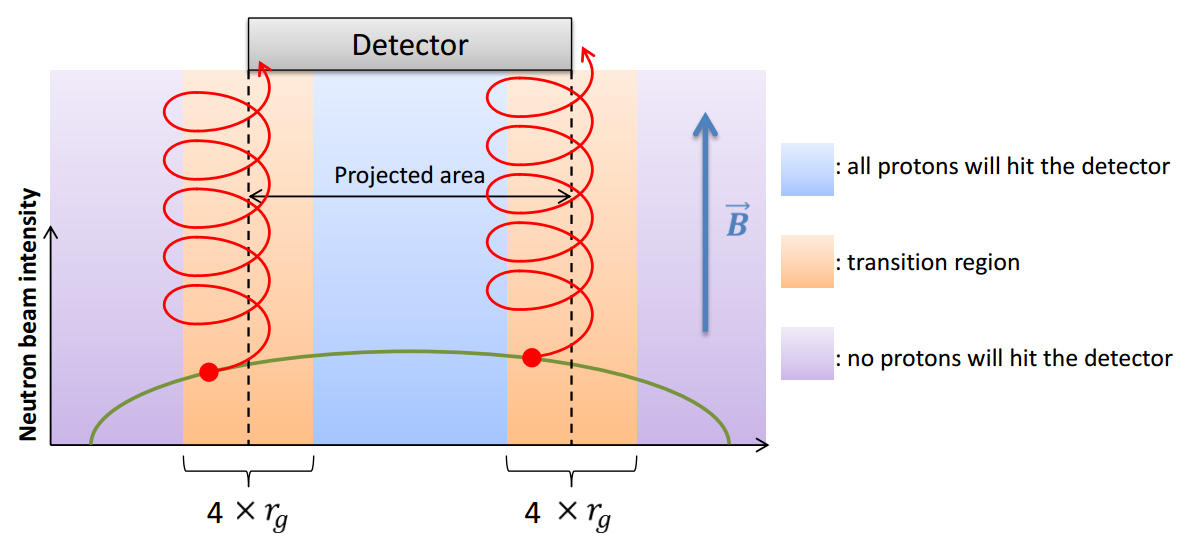}
\end{center}
\caption[Edge effect description.] {Scheme of the edge effect. Proton (in red) on the left is emitted outside of the projected area and hits the detector whereas the one on the right is emitted inside the projected area and misses the detector. Transition regions are centered on the edges of the projected area. Their width is equal to four times the radius of gyration of the concerned particle $r_{g}$.}
\label{EdgeEffect}
\end{figure}

The magnetic field projects the shape of the neutron beam onto the detector: decay protons gyrate around magnetic field lines with a radius of gyration dependent of their energy. Since the neutron beam is wider than the detector, not all of the decay protons can be detected. As shown in Fig. \ref{EdgeEffect}, three different regions can be defined:

\begin{itemize}
	\item Inside the projected area, minus the gyration radius of the particle: in this region, all protons will hit the detector.
	\item The transition region, defined by the edge of the detector plus or minus the gyration radius: depending on their energy, protons emitted inside the projected area might miss the detector, and protons emitted outside the projected area can hit the detector.
	\item Outside the projected area plus the radius of gyration: in this region, no protons can hit the detector.
\end{itemize}

However if the neutron beam profile is homogeneous the two particular cases described for the transition region cancel (assuming a uniform detection efficiency): this is the ideal case. In reality the neutron beam profile is not perfect and therefore, at the edge of the projected area, one effect will be dominant. For example we can see on Fig. \ref{EdgeEffect} that there will be more protons emitted inside the projected area that will miss the detector than protons emitted outside the projected area that will hit the detector. Furthermore, since the radius of gyration depends on the energy of the proton, this effect will have a direct proton energy dependency and thus a direct effect on $a$.

Since the longitudinal neutron beam profile (x-axis) can be considered flat, this effect only concerns the direction transversal to the neutron beam (y-axis). To reduce this effect, one possibility would be to reduce the width of the neutron beam such that all emitted proton would hit the detector. With the actual setup, the impact on the statisctics would not be acceptable. Another solution is to optimize the neutrons aperture in order to get a neutron beam as flat as possible.

This last option was chosen: there is a dedicated neutron guide before the entrance window of $a$SPECT that has been optimised with two different Monte-Carlo simulation softwares (details can be found in \cite{Bor10}). Furthermore a collimation system inside the Decay Volume itself is used to avoid neutrons hitting the spectrometer walls and to refine the collimation\footnote{The collimation inside the spectrometer has to be minimal in order to avoid background radiation.}.

Nevertheless the neutron beam profile is not flat enough to neglect the edge effect. As a result, we use Monte-Carlo simulations in order to quantify it. To check those calculations, we made several measurement of the neutron beam profile. The next section is dedicated to those measurements.

\subsubsection{Copper foil activation}
\label{CopperFoilActivation}

In order to measure the neutron beam profile, we use copper foil activation at different places of the experiment. Those foils are made of natural copper which is composed of roughly 2/3 of $^{63}$Cu and 1/3 of $^{65}$Cu. Under a neutron beam, those two isotopes will capture neutrons to form respectively $^{64}$Cu and $^{66}$Cu which are both beta emitters with a respective half-life of $\tau = 12.700(2)$ h and $\tau = 5.12(14)$ min \cite{Sin07, Bro10}. Approximately one hour after the activation, only the $^{64}$Cu is still really active. By using an X-ray dedicated image plate and scanner, the copper foils can then be read out with a pixel size of $200\,\mu m/pixel$.

\begin{figure}
        \centering
        \begin{subfigure}{0.48\textwidth}
                \centering
                \includegraphics[width=\textwidth]{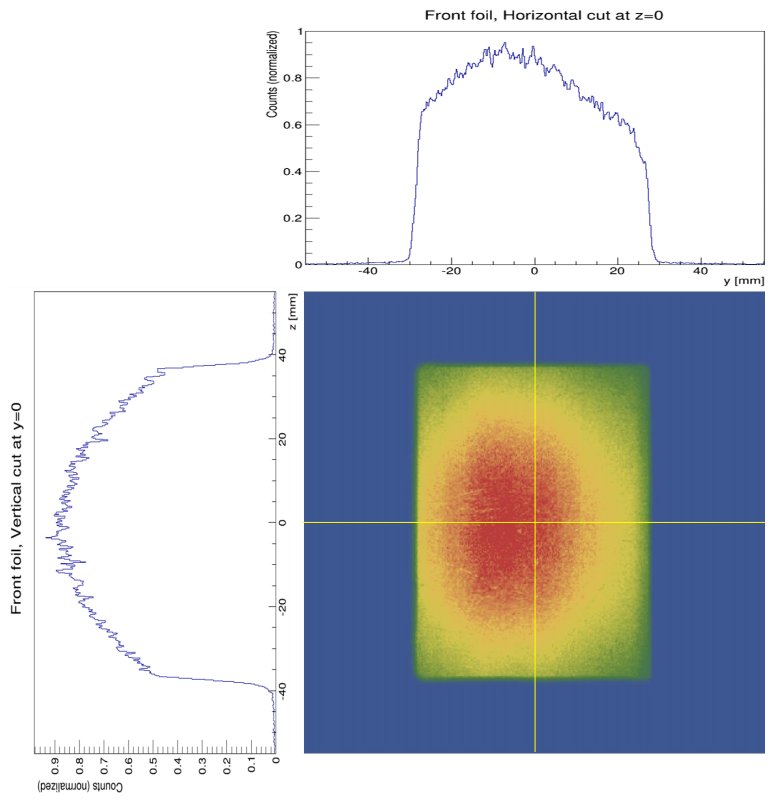}
                \caption{}
                \label{EntranceFoilProfile}
        \end{subfigure}
        \begin{subfigure}{0.48\textwidth}
                \centering
                \includegraphics[width=\textwidth]{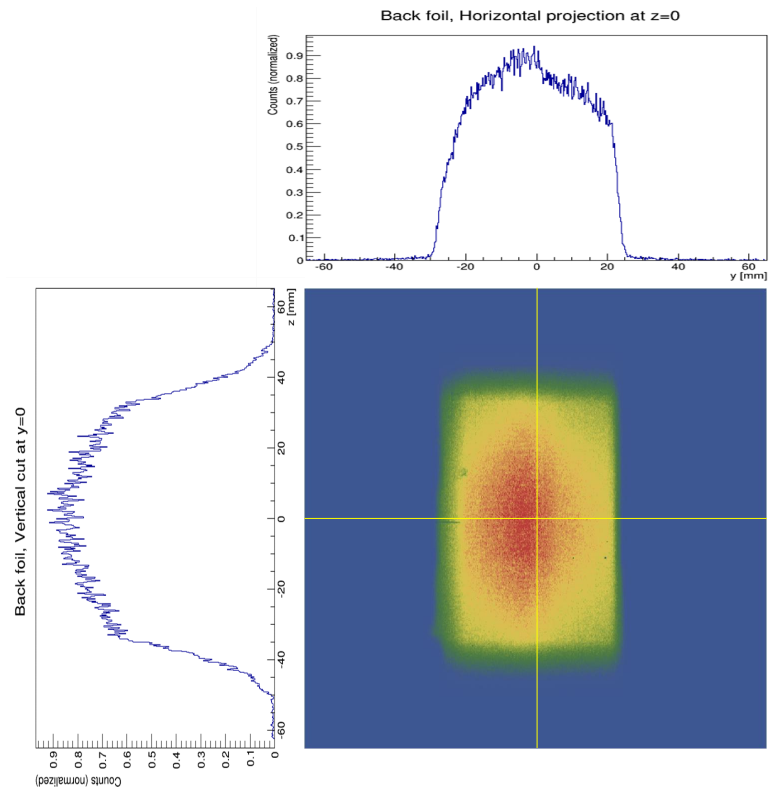}
                \caption{}
                \label{ExitFoilProfile}
        \end{subfigure}
				\vspace{0.5cm}
        \caption[(a): Neutron beam profile at the entrance window of the spectrometer and (b): Neutron beam profile at the exit window of the spectrometer.]{(a): Neutron beam profile at the entrance window of the spectrometer. (b): Neutron beam profile at the exit window of the spectrometer.}\label{WindowBeamProfile}
\end{figure}

At the beginning of the 2013 beam time, we made this measurement with two coppers foils placed respectively in front of the entrance window and behind the exit window of $a$SPECT (Fig. \ref{WindowBeamProfile}). This first acquisition allowed to check the neutron beam profile without any impact on the spectrometer vacuum quality. Indeed the main vacuum has to be severly deteriored if we want to measure the neutron beam profile directly inside the Decay Volume: even-though an airlock permits to move objects inside the decay volume using a manipulator (see Sec. \ref{CopperWireScan}$\,$), the impact on the vacuum quality is non-negligible. As a result, those measurements were done at the very end of the beam time and with two different apertures placed before the entrance window. The first one is 75 mm high and 50 mm wide. It was used during the majority of the beam time and serves to reduce the beam divergence. The second one is narrow (73$\times$15 mm) and thus reduces a lot the neutron beam width (see Fig. \ref{InsideBeamProfile}). We used the narrow aperture for the so-called "narrow beam" $a$ measurements, in order to investigate the edge effect experimentally.

The beam profiles at the entrance and exit of $a$SPECT indicate that the neutron beam is symmetric along the z axis but is shifted by several millimeters toward Chartreuse (y<0). From the profile inside the Decay Volume, with the normal aperture, we denote the same shift.

Concerning the narrow beam profile, it is symmetric in both directions but not centered in x direction. The narrow aperture was intentionally shifted towards Chartreuse in order to obtain the maximum beam intensity at about the same position as with the full beam. This profile and the $a$ measurement realized with this aperture will be used to determine the edge effect systematic effect.

\begin{figure}
        \centering
        \begin{subfigure}{0.48\textwidth}
                \centering
                \includegraphics[width=\textwidth]{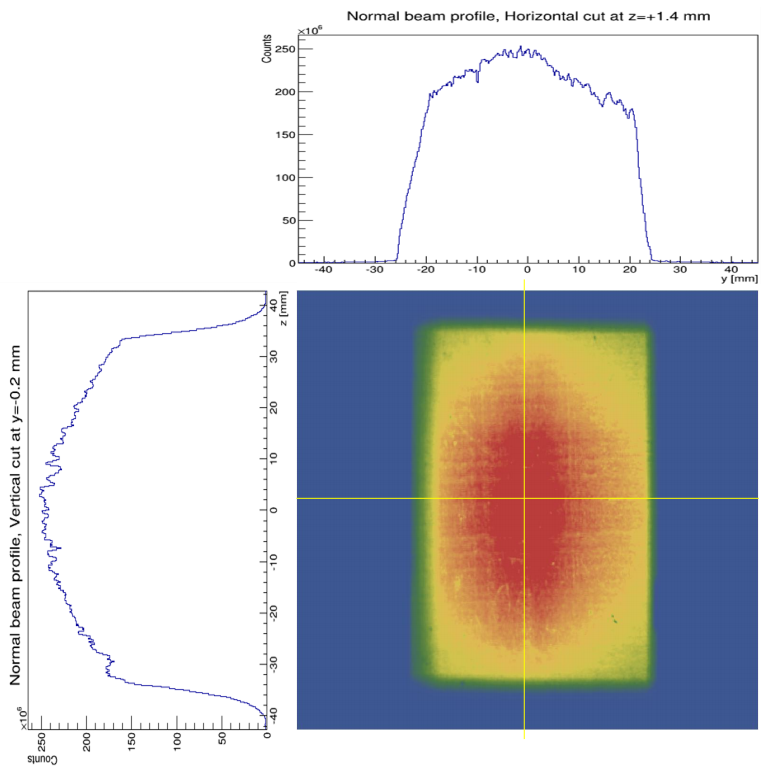}
                \caption{}
                \label{EntranceFoilProfile}
        \end{subfigure}
        \begin{subfigure}{0.48\textwidth}
                \centering
                \includegraphics[width=\textwidth]{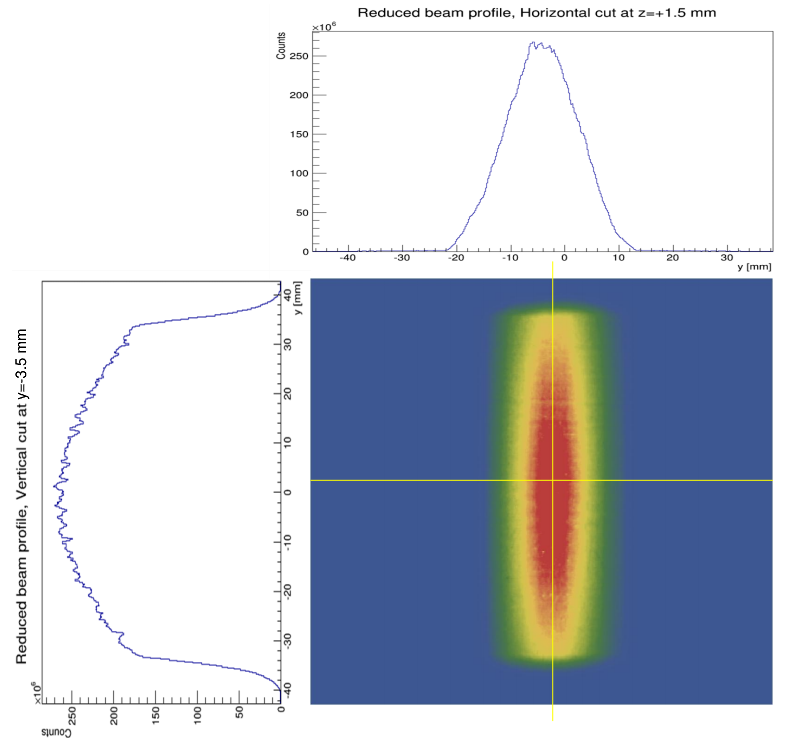}
                \caption{}
                \label{ExitFoilProfile}
        \end{subfigure}
				\vspace{0.5cm}
        \caption[(a): Neutron beam profile inside the spectrometer with normal aperture and (b): Neutron beam profile inside the spectrometer with narrow aperture.]{(a): Neutron beam profile inside the spectrometer with normal aperture. (b): Neutron beam profile inside the spectrometer with narrow aperture.}\label{InsideBeamProfile}
\end{figure}

%%%%%%%%%%%%%%%%%%%%%%%%%%%%%%%%%%%%%%%%%%%%%%%%%%%%%%%%%%%%%%%%%%%%%%%%%%%%%%%
%  New DAQ
%%%%%%%%%%%%%%%%%%%%%%%%%%%%%%%%%%%%%%%%%%%%%%%%%%%%%%%%%%%%%%%%%%%%%%%%%%%%%%%

\clearpage

\section{The new data acquisition system (new DAQ)}
\label{newDAQ}

Since the beginning of the $a$SPECT experiment, several upgrades of the electronics have been realized (see Sec. \ref{OldDAQSignalProcElec} ). However, some problems subsist with this electronic setup. First of all, the used ADC was an in-house development of the group E18 at the Technical University Munich for the COMPASS experiment. No support is available anymore. Therefore this system relies on a single old computer with very specific hard- and software. This computer relies
on other computers that handle logfiles, electrodes voltages, electrodes currents, data writing etc.. This intricacy of computers and communications make it a very complex and very sensitive system and all this setup is getting quite old, resulting in occasionnal crashes. From a physics point of view, this data acquisition system (DAQ) also presents some important problems, such as a big temperature dependency of the shaper or a problematic behavior after high-energy electrons (those issues will be explained in detail in the following sections). Finally, all the data acquisition Labview softwares have been developped by a single person only, Dr. Martin \textsc{Simson}, making him the only expert of this installation. Since his contract with the ILL will be finished soon, it is important for the future of the experiment to have a long time support for the DAQ.

\begin{figure}[h!]
\begin{center}
\includegraphics[width=110mm]{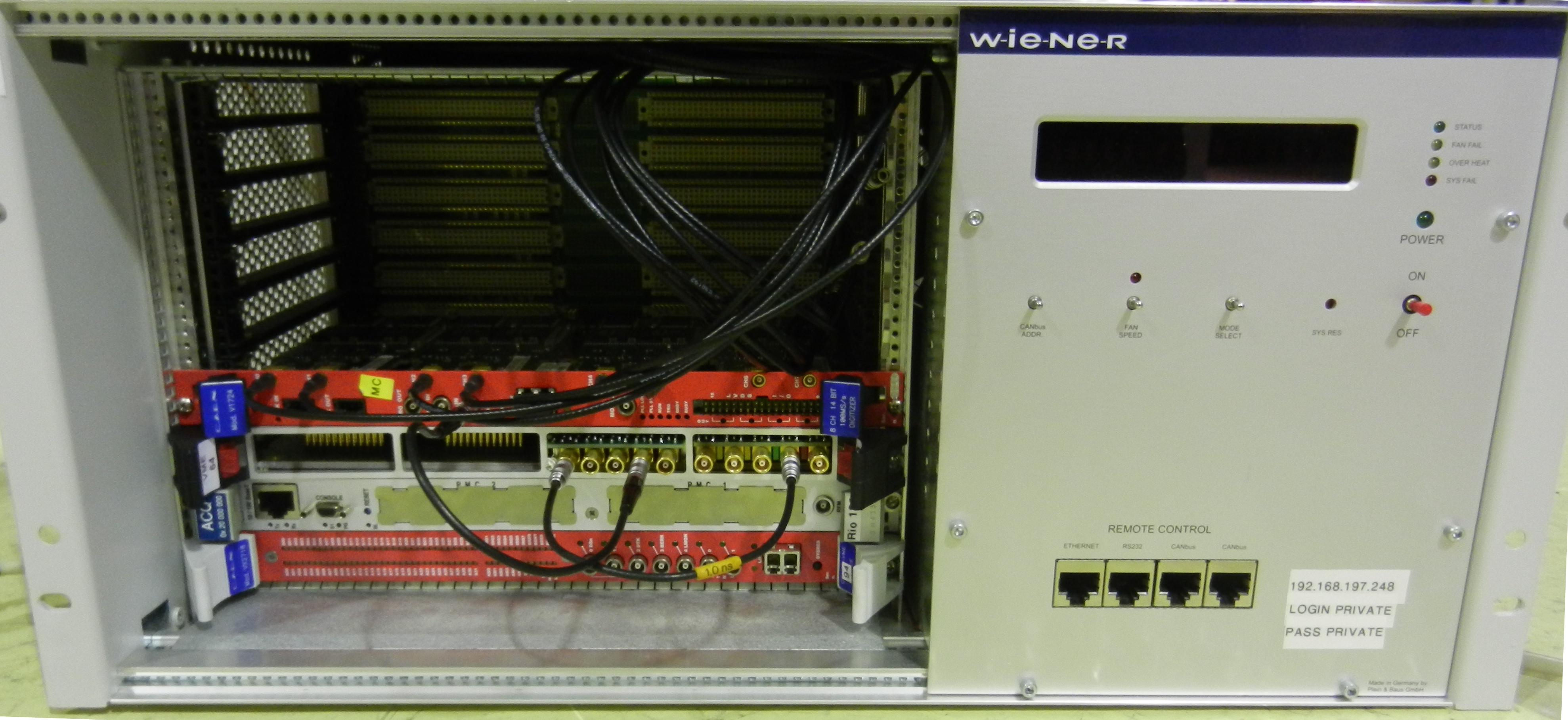}
\end{center}
\caption[New DAQ crate.] {Picture of the new DAQ crate.}
\label{DAQCrate}
\end{figure}

The ILL possesses a dedicated data acquisition group of about 15 people: the SCI (Service de Controle des Instruments). This team has developed their own instrument control software called NOMAD. It is written in C++ and Java and they can adapt it to an experiment. As a result they provided us a new DAQ (Fig. \ref{DAQCrate}) and a new acquisition interface. 

\subsection{Description of the new DAQ}

The new acquisition system  is built around a commercial Analog to Digital Converter (ADC): a V1724 card made by CAEN\footnote{\textbf{C}ostruzioni \textbf{A}pparecchiature \textbf{E}lettroniche \textbf{N}ucleari. This company is specialized in "the design, the production and the supply of electronic instrumentation for radiation and low light sensors".}. Over the last few years flash ADC\footnote{A flash ADC compares the input voltage with a certain number of reference voltages. All those comparisons are made at the same time, in parallel, resulting in a very high sampling rate. However the better the resolution the more comparators you need: $2^{n}-1$ for a $n$-bit conversion.} have greatly evolved, allowing to get higher resolution and higher sampling frequency for a reasonable price. This V1724 is equipped with an 8 channels, 14 bit and 100MHz sampling frequency flash ADC. There are also several FPGAs dedicated to the energy calculation of an event (more details on Sec. \ref{EnergyCalculation}) directly implemented on this card.

The main difference compared to the old DAQ is the high resolution (14 bit instead of 12 bit). Therefore the full dynamic range of the preamplifier can be covered still providing sufficient resolution for the tiny proton signal. The shaper is no longer required: the digitalization is made as soon as possible, just after the preamplifier. Once digitized the signal is directly treated by the algorithms programmed into the FPGAs of the board. Digital signal treatment possesses several advantages over an analog signal treatment, such as a great reproducibility and a great flexibility through FPGAs programmation.

\begin{figure}[h!]
\begin{center}
\includegraphics[width=130mm]{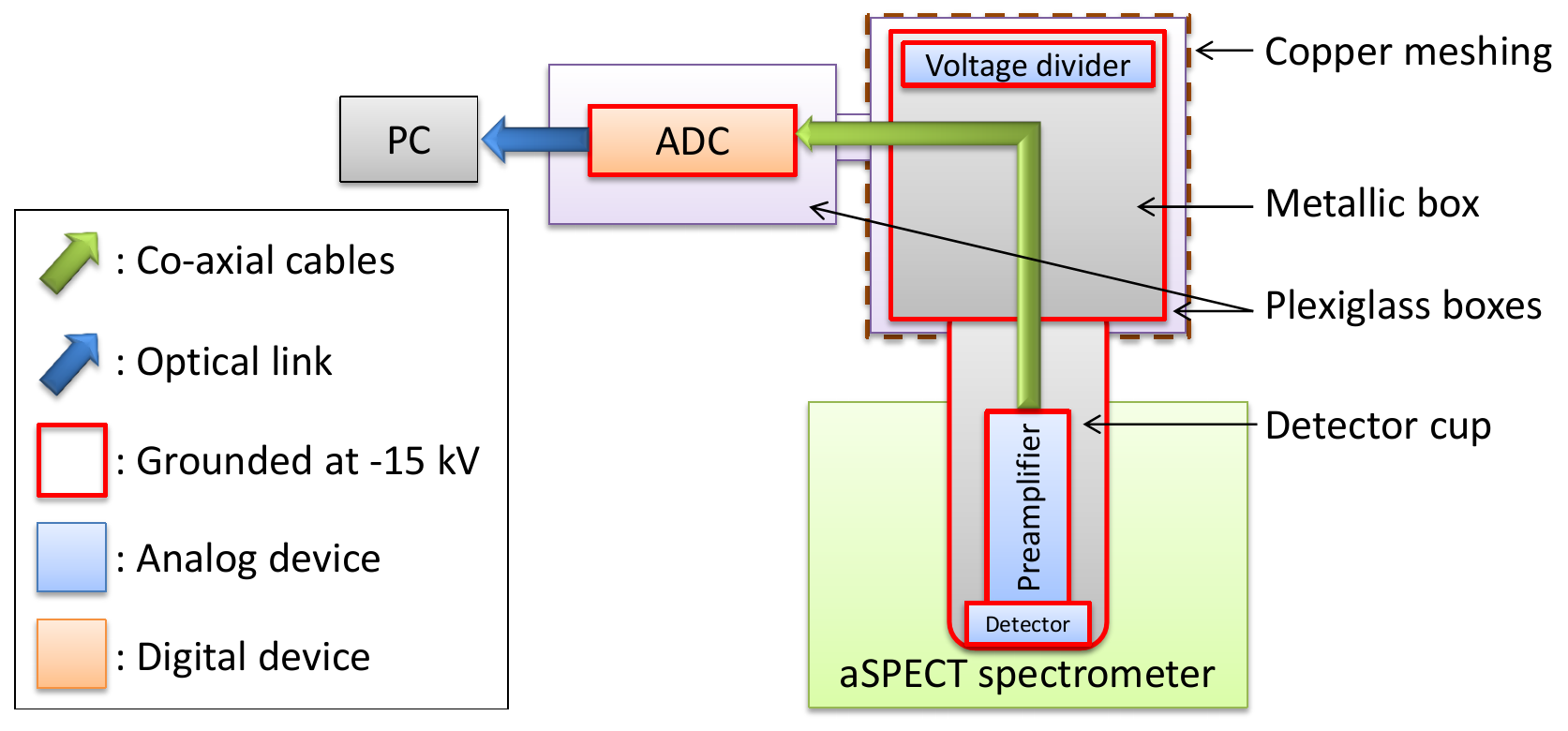}
\end{center}
\caption[Scheme of the new DAQ installation on $a$SPECT.] {Scheme of the new DAQ installation on $a$SPECT.}
\label{NewDaqScheme}
\end{figure}

Since the new DAQ crate is too big to fit the plexiglass box at the top of $a$SPECT, it is placed inside another one just aside (Fig. \ref{NewDaqScheme}). This second box already contains the power transformer (used to provides 220V AC current uncoupled from the Electricite de France' ground) and the +30/-30 V alimentation that supplies the detector and its electronics.

\subsubsection{Energy calculation}
\label{EnergyCalculation}

\begin{figure}[h!]
\begin{center}
\includegraphics[width=120mm]{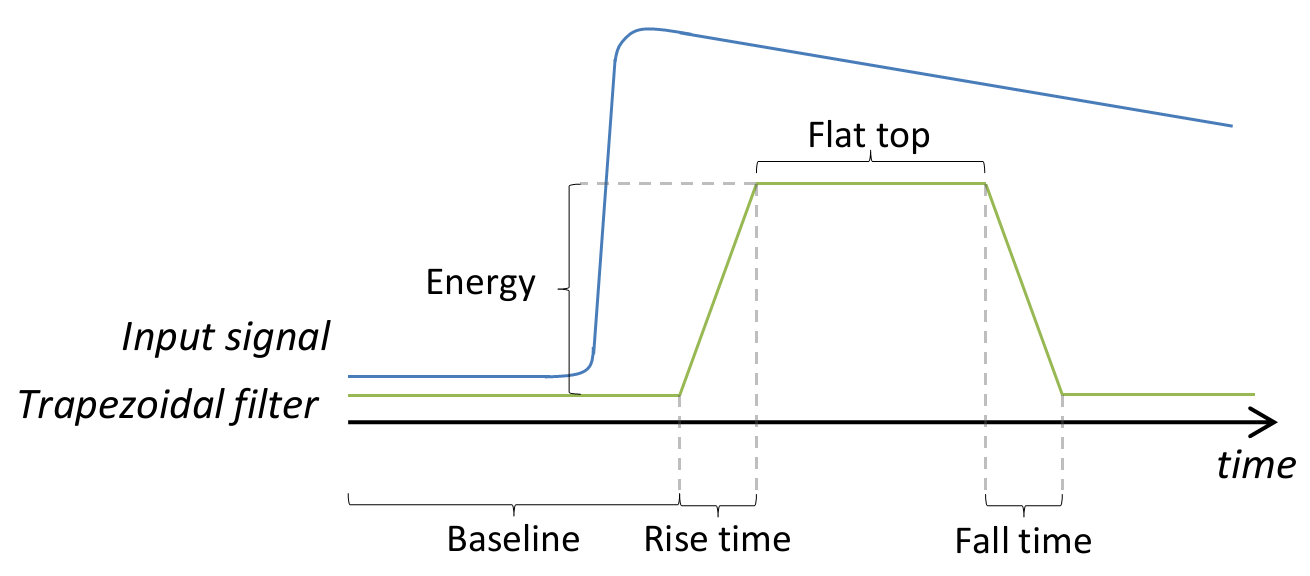}
\end{center}
\caption[Energy calculation with the new DAQ.] {Energy calculation with the new DAQ. The baseline value is calculated by averaging a certain number of trapezoidal filter samples before the pulse. The rise and fall time are always the same. The flat top level allows to determine the energy by subtracting the baseline value from it. The longer the sampling time (trapezoid length), the better the energy resolution but also the pile-up probability.}
\label{Energy}
\end{figure}

In order to determine the energy deposited by a particle in the detector, a trapezoidal filter is applied on the input signal directly at the V1724 board. This digital filter transforms the typical exponential decay signal into a trapezoid, as shown in Fig. \ref{Energy}. 

The energy is directly proportionnal to the height of the flat top of the trapezoid minus the baseline value and is given as a 15 bit integer. All those parameters (baseline length, rise time of the trapezoid, flat top length etc.) can be defined in NOMAD which serves as interface.

\subsubsection{Trigger condition}
The counterpart of a very high sampling frequency is a substantial flow of data: more than 140 MB/s per channel for a $10 \, $ns time resolution. In order to treat only the relevent data, a digital pulse triggering filter analyses continuously the input signals. To efficiently discriminate noise and not to loose any event, this trigger algorithm is a digital version of an analog RC-(CR)$^{2}$ filter. The RC filter (also called \textit{integration filter}) is a low-pass filter and thus suppresses high-frequency noise whereas the CR filter (also called \textit{derivative filter}) is a high-pass one and suppresses low-frequency noise such as ground loops. It also applies a baseline restorer, resulting in a very stable baseline for the trigger. 

\begin{figure}[h!]
\begin{center}
\includegraphics[width=110mm]{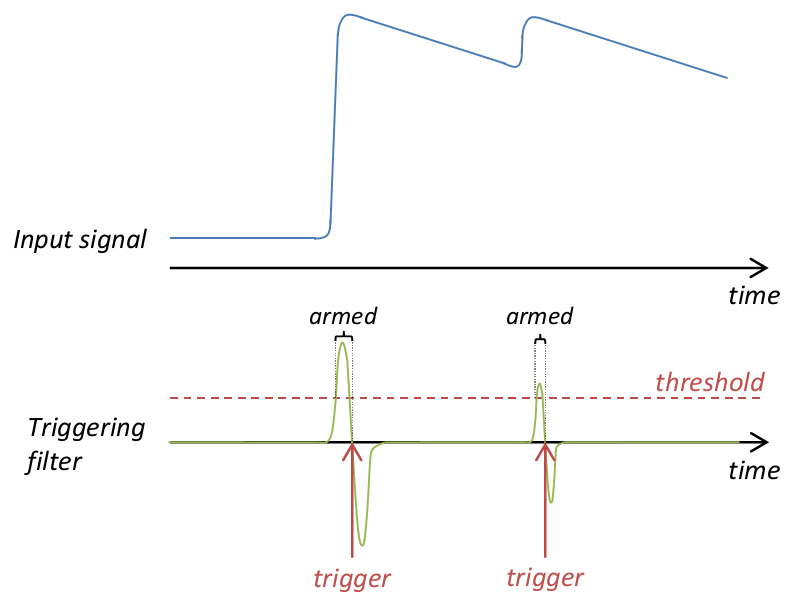}
\end{center}
\caption[Trigger description.] {Trigger exemple for two consecutive events. When the triggering filter signal exceeds the threshold, the trigger is armed and will only be sent when a zero-crossing occurs. However if the trigger is armed and no zero-crossing happens before a defined delay, no trigger will be sent.}
\label{Triggering}
\end{figure}

When treated by this filter a pulse will be transformed into a bipolar signal, as shown in Fig. \ref{Triggering}. When the signal amplitude goes above the threshold value the trigger is armed and it is only when the baseline ("zero") is crossed again that the trigger signal is sent: it is the timestamp of the triggered event. As a result the definition of the timestamp of an event is not dependent of the pulse amplitude.

\begin{figure}[h!]
\begin{center}
\includegraphics[width=90mm]{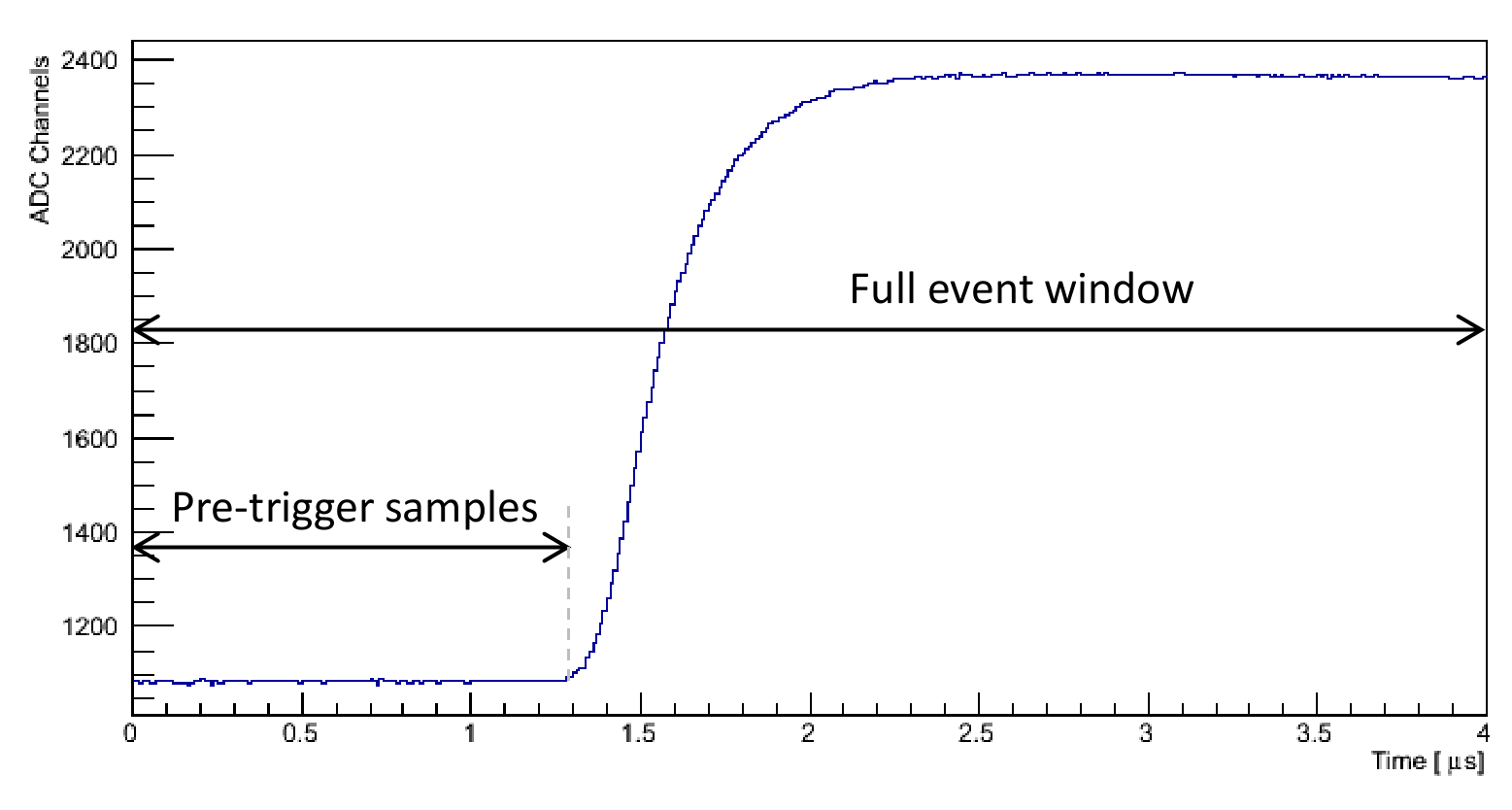}
\end{center}
\caption[New DAQ event waveform example.] {Example of an event waveform with the new DAQ.}
\label{EventMusec}
\end{figure}

Once a trigger is sent the event is saved and the corresponding window is defined by the total number of samples and the number of samples before the trigger, as shown in Fig. \ref{EventMusec}. During this window no other events can be saved\footnote{An online pile-up rejection system is available on the new DAQ but his behavior is not reliable: for unknown reasons the timestamp is lost when a pile-up occurs with this option activated. As a result it was not used.} and piled-up events that occur in the event window can be treated with an offline analysis (more details on such analysis,
performed for the old DAQ, can be found in \cite{Sim10}).

\subsubsection{Dead time}

If we don't consider any offline analysis for pile-up events, the dead time of the DAQ is the duration of the event window plus the electronics latency. This time is the minimal delay between two consecutive events on the same channel. To correct this loss of events we need to apply a dead time correction expressed as:

\begin{eqnarray}
C_{{\rm corr}}=\frac{C_{{\rm meas}}}{1-C_{{\rm total}}\times T_{{\rm dead}}}
\end{eqnarray}

with $C_{{\rm total}}$ the total trigger rate, $C_{{\rm meas}}$ the measured count rate for the region of interest (like protons or electrons) and $C_{{\rm corr}}$ the corrected count rate for the region of interest. $T_{{\rm dead}}$ is the dead time per event.

It is important to know the dead time value precisely in order to apply a good correction: an uncertainty on this parameter results in a systematic error on the $a$ value in two different ways: the total trigger rate and thus the (uncorrelated) dead time correction depends on the analysing plane voltage $U_{\rm A}$ and protons can be detected correlated to electrons. In the $a$SPECT experiment the minimum time difference between an electron hitting the detector and the correlated proton is $5.2\,{\rm \mu s}$ \cite{Kon11}. As a result a dead time higher than this would involve possible loss of correlated proton events. 

The minimum time difference between two consecutives events of a same channel for the old DAQ has been measured as $4.2 \, \mu s$. Since the event window is $4\,{\rm \mu s}$ long, there is a $0.2 \, {\rm \mu s}$ latency due to the electronics. For the new DAQ a $4.06 \, {\rm \mu s}$ dead time value has been measured, resulting in a $0.06 \, {\rm \mu s}$ electronic latency. The precision on both measurements is the time resolution which is respectively $0.05 \, {\rm \mu s}$ and $0.01 \, {\rm \mu s}$ for the old and the new DAQ.

\subsubsection{Data structure}

\begin{figure}[h!]
\begin{center}
\includegraphics[width=130mm]{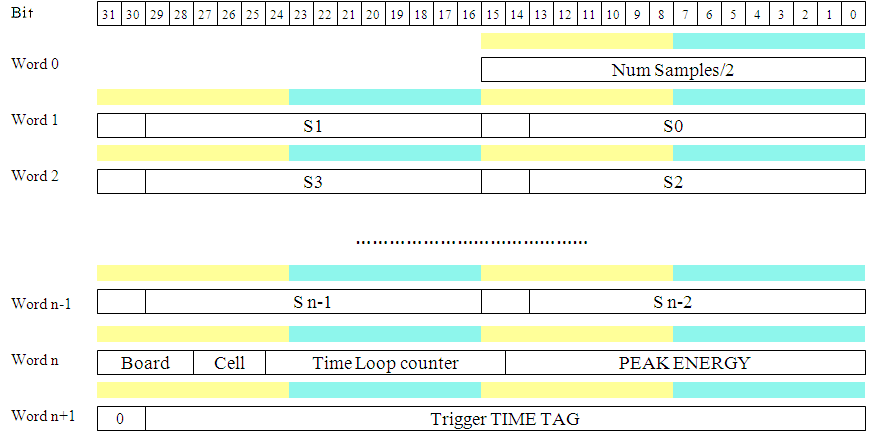}
\end{center}
\caption[List mode data structure.] {List mode data structure of the NOMAD output binary file.}
\label{list_mode_format_aSPECT}
\end{figure}

The acquisition output from NOMAD is a "list mode" binary file (extension .lst) with the structure indicated in Fig. \ref{list_mode_format_aSPECT}. It is a succession of 32 bit words, with the 4 first words of a file being "dummy words" and thus ignored. For a $4\,{\rm \mu s}$ event length, each event will be composed of 203 words. The first one is the number of ADC slices divided by 2. Since the time resolution is $10\,{\rm ns}$ all events contains 400 ADC slices, meaning that the first word will always be 200. The next 200 words are those ADC slices, with a 14-bit resolution. Finally the two last words contain the board number (in case of multiple ADC cards), the channel number of the event, the energy calculated from the trapezoid filter, the time interval (called TimeLoop counter in Fig. \ref{list_mode_format_aSPECT}) and the timestamp of the event.

The timestamp has a 30 bit resolution. Therefore, after $\approx 10.74\,{\rm s}$ ($10\,{\rm ns}\times(2^{30})$) of acquisition this value will be resetted to 0 and the time interval incremented by 1. As a result the time in seconds of an event can be calculated as: $t_{{\rm in\, sec}}=({\rm timestamp}+(2^{30})\times {\rm time\,interval})\times 1\times 10^{-8}$.

To analyse the data efficiently those binary files are converted into ROOT files using a C++/ROOT macro specifically developped for this purpose. In order to get the best compatibility with the previous analysis routines developped for the old DAQ, they have a similar structure.

An $a$SPECT ROOT files explorer software has also been created using the GUI (Graphical User Interface) possibilities of the ROOT library. This program allows to quickly check the data by looking at the events, the resulting energy spectrum or the events time distribution for example. It is possible, for every of those plots, to apply channel filters and cuts on several variables (such as the time, energy, baseline value etc.).

\begin{figure}
\begin{center}
\includegraphics[width=160mm]{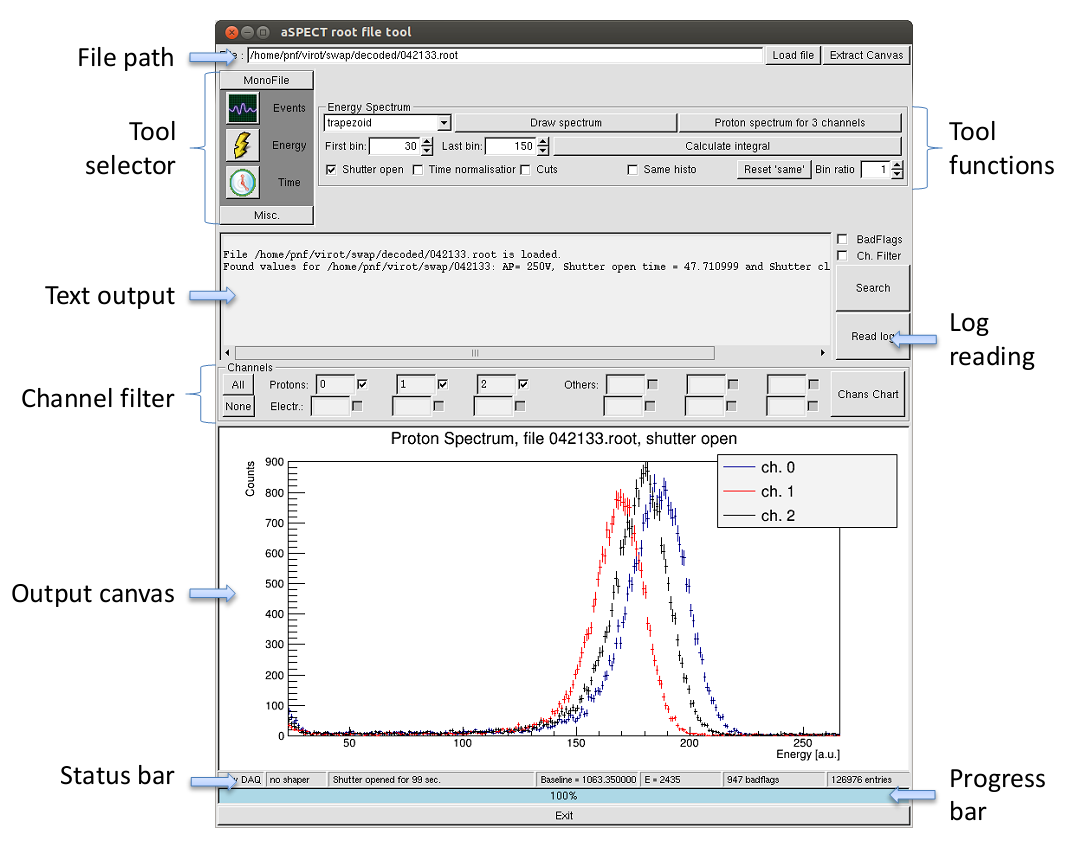}
\end{center}
\caption[$a$SPECT "Tool" software.] {$a$SPECT "Tool" software screenshot.}
\label{Gui}
\end{figure}

\subsection{Online behavior}
After testing the new DAQ outside the spectrometer with a small test setup and $\beta / \gamma$ sources, it was installed on the $a$SPECT experiment for in-situation tests. This few days long data set allowed to directly compare both DAQs in $a$ measurement conditions. Table \ref{DetChansPads} indicates the correspondance between DAQ channels and detector pads.

\begin{table}[h!]
	\centering
		\begin{tabular}{|c|c|c|}
			\hline
			\textbf{Pad} & \textbf{Old DAQ Channel} & \textbf{New DAQ Channel} \\
			\hline
			1 & 21 & 0\\
			\hline
			2 & 20 & 1\\
			\hline
			3 & 19 & 2\\
			\hline
		\end{tabular}
		\vspace{0.5cm}
	\caption{Relation between detector pads and both DAQs channels.}
	\label{DetChansPads}
\end{table}

%\subsubsection{Energy resolution and noise separation}

%The energy resolution of the two DAQs can be compared by measuring the proton peak width and then normalize it in regards of the peak position. Fig. FIGURE represents the energy spectrum of each DAQ for the same detector high-voltage ($-15$ kV) and similar experiment conditions. The proton peak is fitted, as a first order, with a gaussian.

\subsubsection{Proton peak stability}

\begin{figure}[h!]
\begin{center}
\includegraphics[width=110mm]{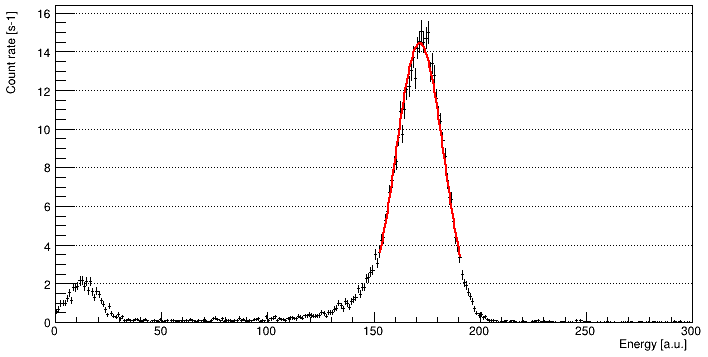}
\end{center}
\caption[Proton peak fitted with a gaussian.] {Proton peak fitted with a gaussian to determine its position.}
\label{ProtonFitgauss}
\end{figure}

Evolution of the proton peak energy position is an indicator of the system stability. Even though it depends on the detector, the preamplifier and the acquisition system, a comparison between the two configurations will indicate which DAQ has the worst contribution. In order to do so the position of the proton peak has been fitted with a gaussian for each files of the selected data set (see Fig. \ref{ProtonFitgauss} for example).

\paragraph{Old DAQ}

\begin{figure}[h!]
\begin{center}
\includegraphics[width=100mm]{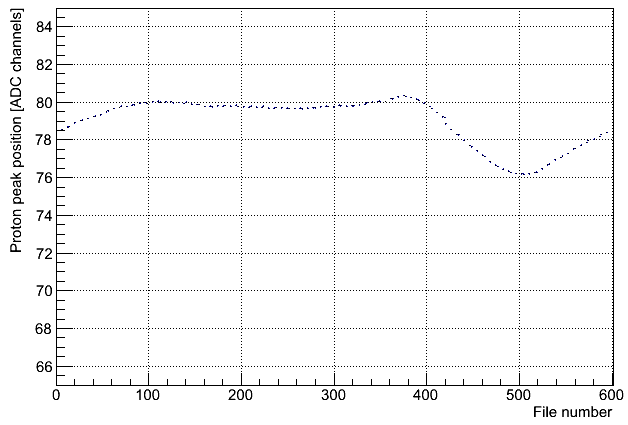}
\end{center}
\caption[Proton peak position, central pad, 2013/06/28.] {Evolution of the proton peak position for the central pad (channel 20) and for AP=50V. This data set regroups three days of measurements, from 2013/06/28 to 2013/06/30.}
\label{PeakPosOld}
\end{figure}

The main source of amplification instability for the old DAQ comes from the shaper and its analog components. Those are quite sensitive to the temperature and the amplification can change depending of the season, weather, period of the day etc.. In order to reduce this effect and to cool the electronics, a compressed air tube constantly blows air in the metallic box containing the shaper. However, as shown in Fig. \ref{PeakPosOld} the variations are quite important: in the order of $2.5 \, \%$ for this data set.

\begin{figure}[h!]
\begin{center}
\includegraphics[width=100mm]{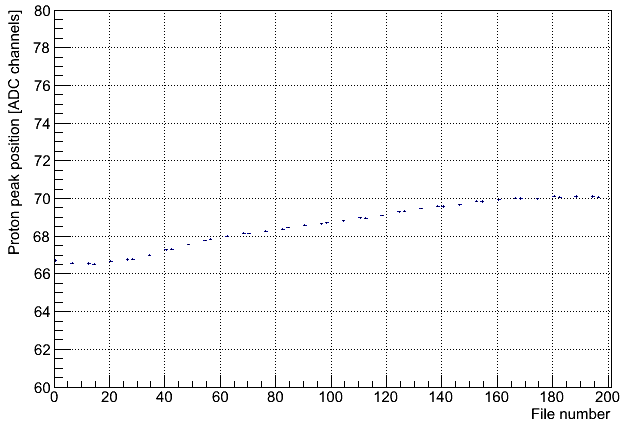}
\end{center}
\caption[Proton peak position after cooling device installation, central pad, 2013/07/25.] {Evolution of the proton peak position for the central pad (channel 20) and for AP=50V after the installation of the cooling device for the air flow. Data acquired the 2013/07/25.}
\label{PeakPosOldAfter}
\end{figure}

A regulated cooling device has been used for the air flow to stabilize more efficiently the shaper temperature. As the compress air tube have to go through the machine, it had to be extended. Because of the longer tube required by this machine the air flow was reduced and the temperature of the shaper increased, resulting in a $\approx13\,\%$ worst amplification (Fig. \ref{PeakPosOldAfter}). Furthermore the absolute variation did not really changed and the relative one thus increased to approximately $3 \, \%$.

\paragraph{New DAQ}

The absence of the shaper with the new DAQ setup implies that the amplification is only dependent of the detector, the preamplifier and the digital trapezoidal filter. 

\begin{figure}[h!]
\begin{center}
\includegraphics[width=110mm]{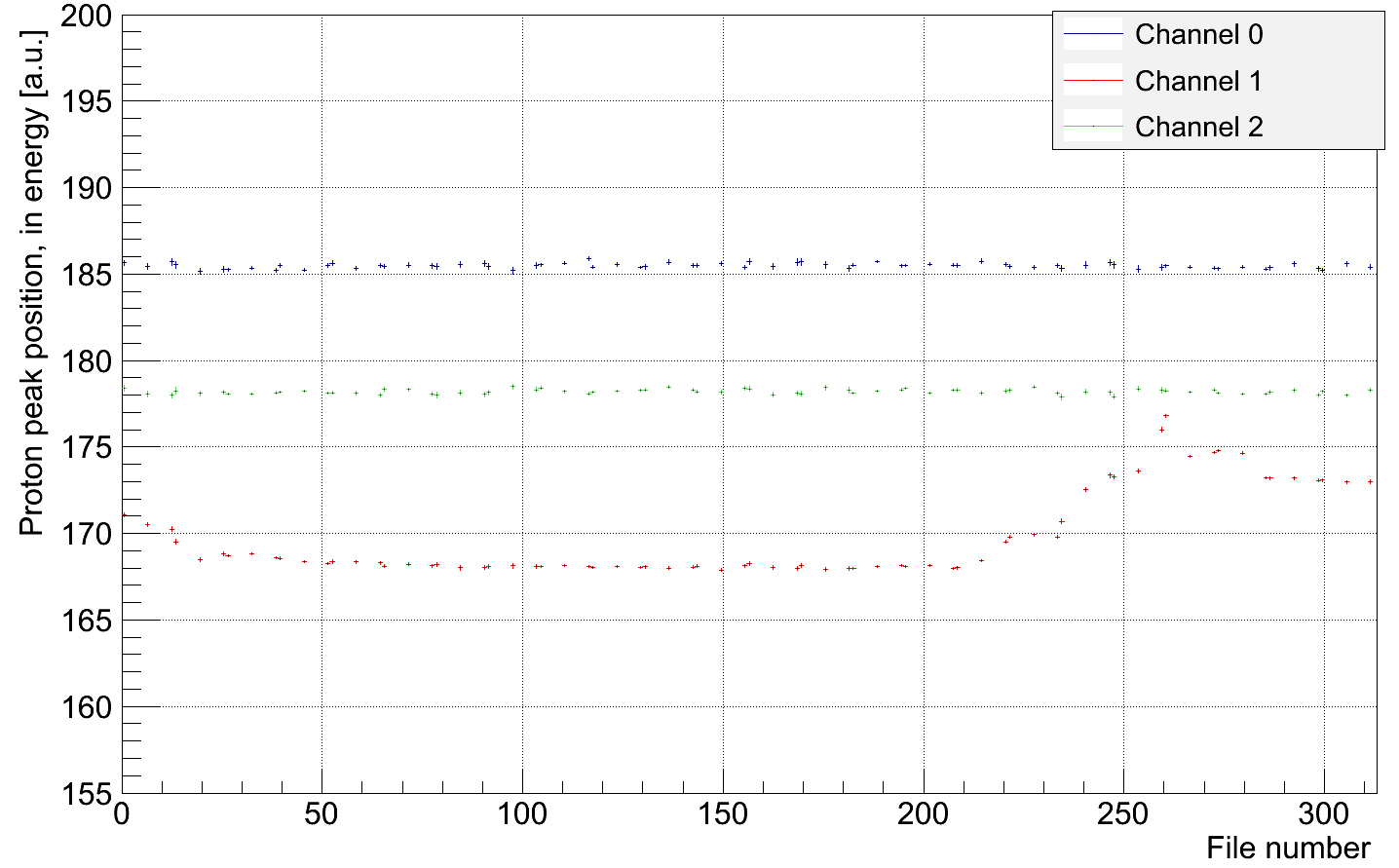}
\end{center}
\caption[Proton peak position with the new DAQ, 2013/07/19.] {Evolution of the proton peak position for all pads with AP=50. Data acquired the 2013/07/19.}
\label{NewDaqPos1}
\end{figure}

Fig. \ref{NewDaqPos1} is the plot of the proton peak position over the 2013/07/19. For channels 0 and 2, the position is extremely stable (less than $0.6\,\%$ variation) but for the central pad we denote a strange behaviour. The proton peak position varies a lot and by steps. Because this problem only concerns channel 1 we can assume it is not a temperature effect (all 3 channels are on the same card and close to each other).

\begin{figure}[h!]
\begin{center}
\includegraphics[width=110mm]{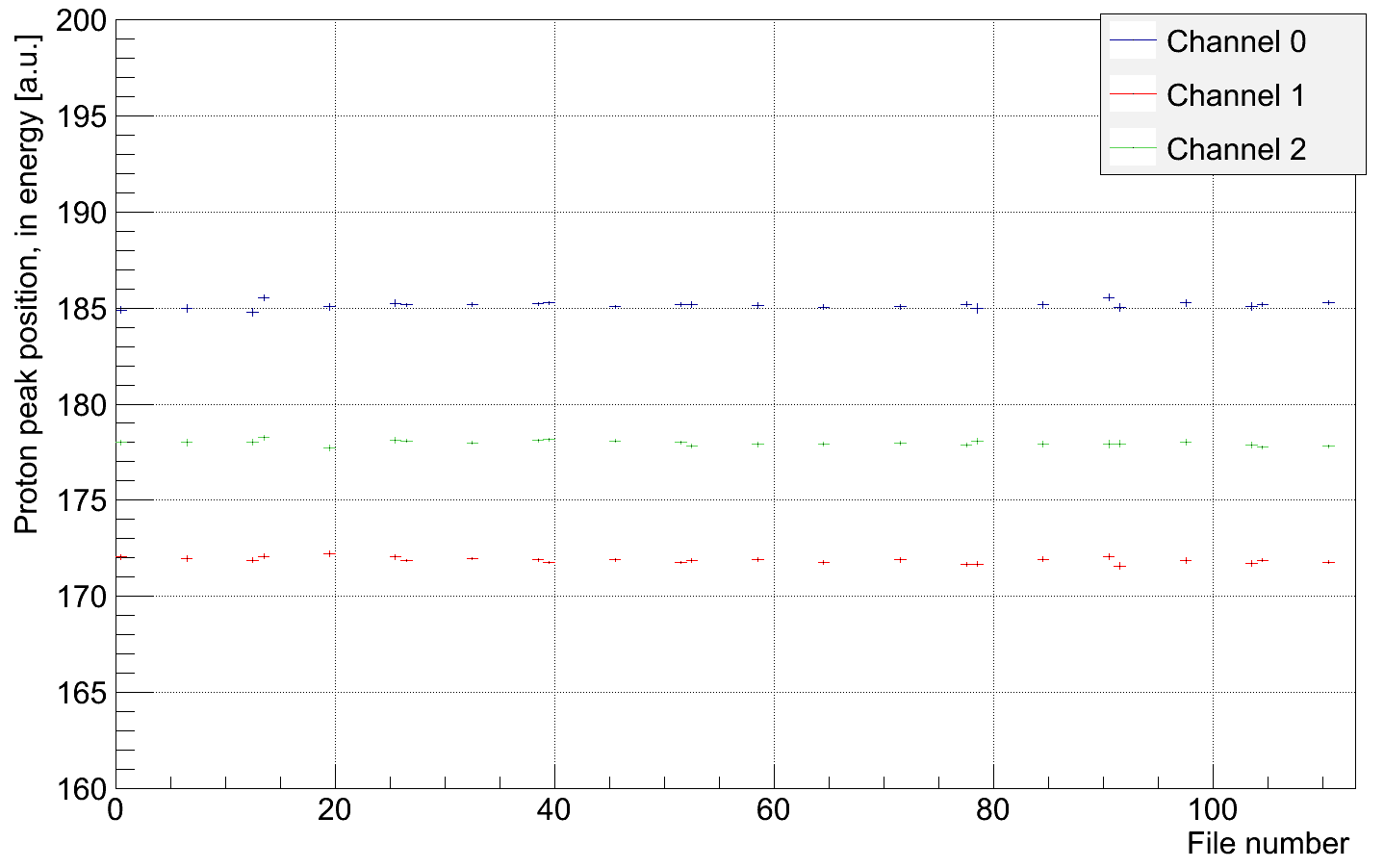}
\end{center}
\caption[Proton peak position with the new DAQ, 2013/07/21.] {Evolution of the proton peak position for all pads with AP=50. Data acquired the 2013/07/21}
\label{NewDaqPos2}
\end{figure}

Furthermore, Fig. \ref{NewDaqPos2} shows the same analysis for the data set of 2013/07/21. All three channels are similar and stable. Those results do not give a clear hint to the cause of the problem but it can be related to:

\begin{itemize}
	\item The detector itself, because of some voltages instabilties from the voltage divider board for example.
	\item The first level of amplification of the preamplifier, which is separate for each channel.
	\item The channel 1 of the DAQ itself. Very easy to fix: just use another channel, but the detector high-voltage have to be ramped down to do so.
\end{itemize}

Because such a behavior has not been seen with the old DAQ, the last option would be the most probable. Since this problem only appeared for two measurements and then vanished, we cannot conclude anything and more in-situation measurements with the new DAQ are required to check this parameter.

\subsubsection{Behavior after high-energy events}

One of the biggest challenge for the data acquisition of $a$SPECT is to handle correctly low-energy proton events that occurs after high-energy electrons. The maximum proton energy after the post-acceleration is $\approx 15$ keV and around $780$ keV for electrons. Fig. \ref{InsideBeamProfile} shows the waveform of such events.
We need to ensure that those protons events are not lost (because of bad trigger conditions for example) and that their energy is not miscalculated because of the shifted baseline due to the previous high-energy event.

\begin{figure}[h!]
        \centering
        \begin{subfigure}{0.48\textwidth}
                \centering
                \includegraphics[width=\textwidth]{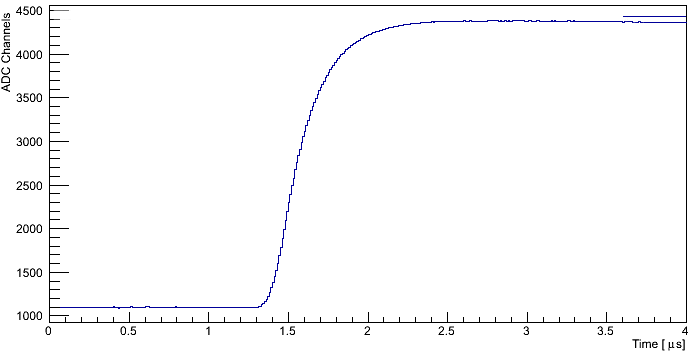}
                \caption{}
                \label{NewDaqElectronTimeDiff}
        \end{subfigure}
        \begin{subfigure}{0.48\textwidth}
                \centering
                \includegraphics[width=\textwidth]{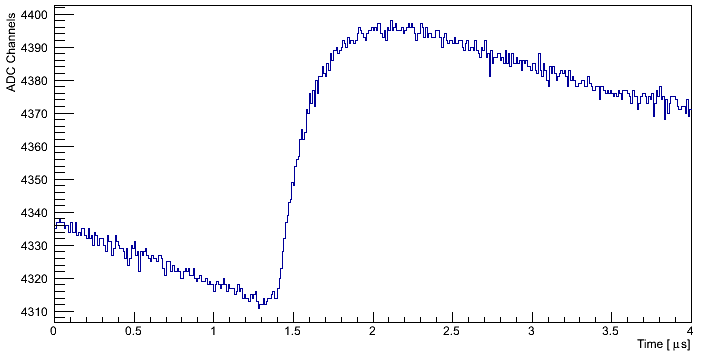}
                \caption{}
                \label{ProtonCorrelated}
        \end{subfigure}
				\vspace{0.5cm}
        \caption[(a): Electron event waveform with the new DAQ and (b): Proton event waveform after a high energy electron.]{(a): Waveform of a high energy electron. The blue bar on the top right represent the scaled height of the correlated proton event (Fig \ref{ProtonCorrelated}). (b): Correlated proton event on the high-energy electron tail. The time difference between those two events is $5.32\,{\rm \mu s}$.}\label{InsideBeamProfile}
\end{figure}

\paragraph{Old DAQ}

\begin{figure}[h!]
\begin{center}
\includegraphics[width=100mm]{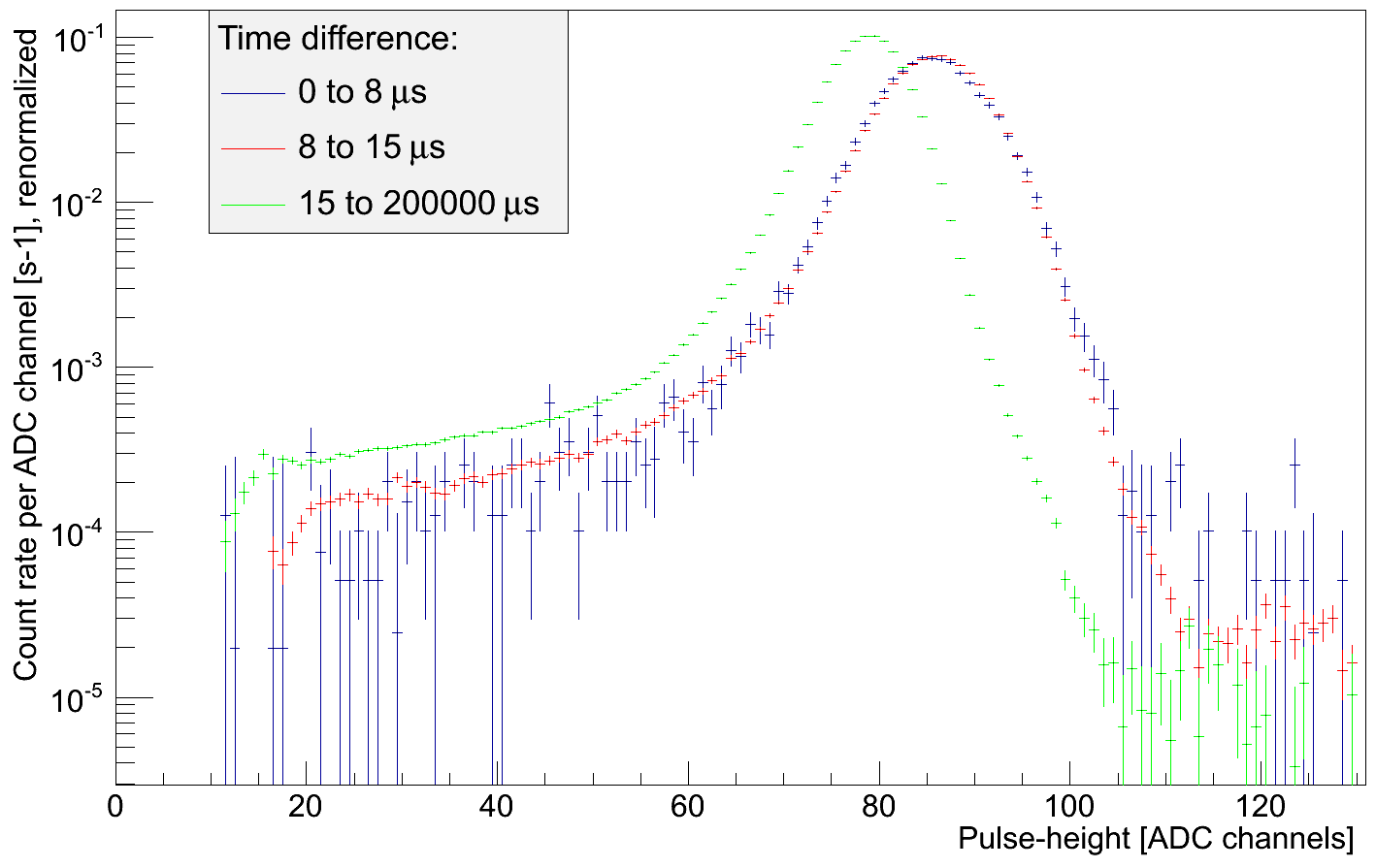}
\end{center}
\caption[Time difference spectrum after high-energy electron, channel 20, old DAQ.] {Time difference spectrum after high-energy electron for the channel 20 and with the old DAQ.}
\label{TimeDiffOldDaq}
\end{figure}

An efficient way to check the good behavior of the DAQ is to look at the proton spectrum depending of the time difference between protons and high-energy electrons. Fig. \ref{TimeDiffOldDaq} shows the result of such a test with the old DAQ: for each channel protons events are sorted depending of their delay with the previous high-energy event. The curves are then normalized in regard of the proton peak region.

\begin{figure}[h!]
\begin{center}
\includegraphics[width=120mm]{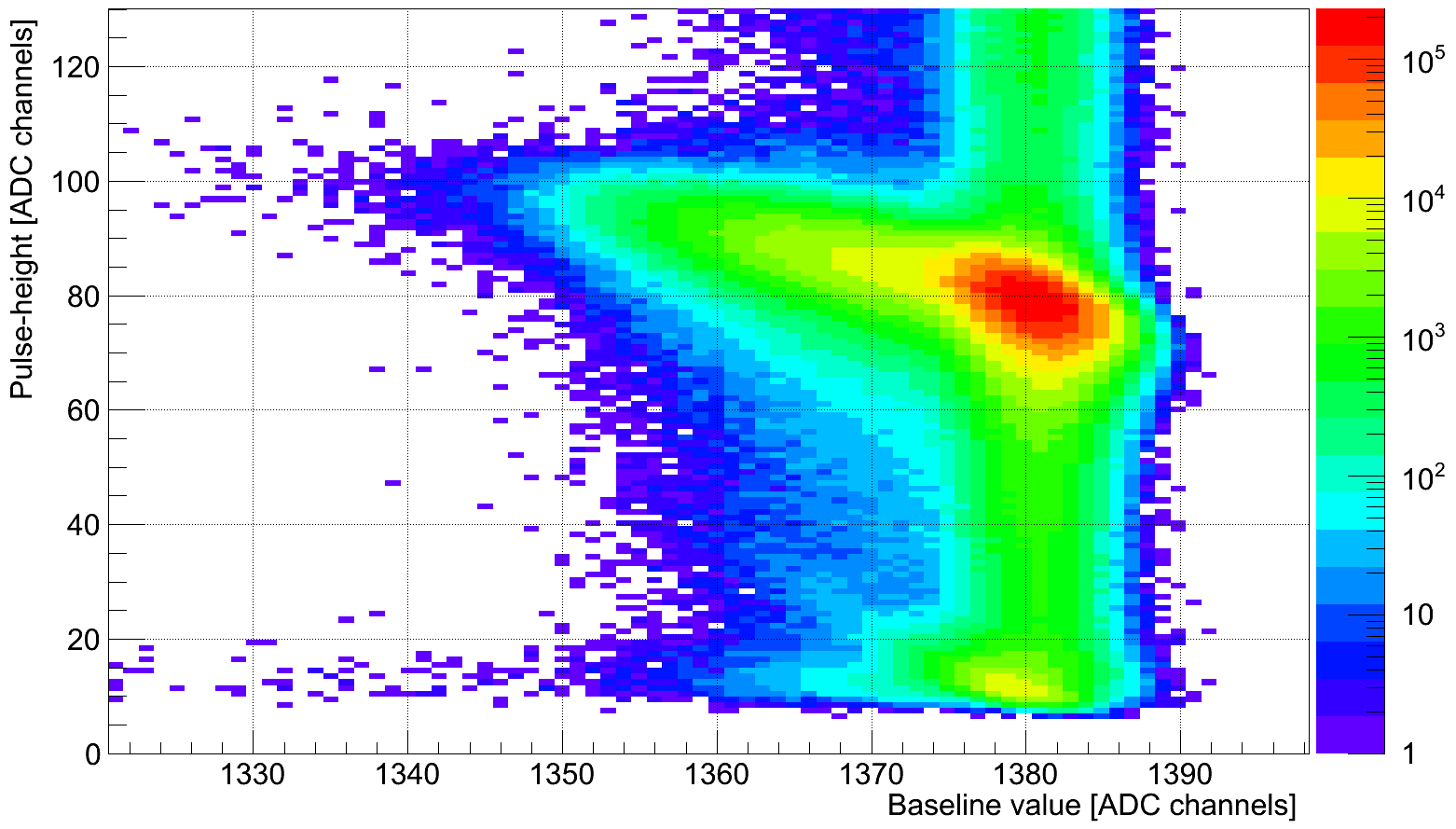}
\end{center}
\caption[Pulse-height VS baseline value with the old DAQ, channel 20.] {Pulse-height VS baseline value with the old DAQ, channel 20.}
\label{TimeDiffOldDaq2D}
\end{figure}

We can see a clear shift in the calculated energy for protons hitting the detector less than $15\,{\rm \mu s}$ after a high-energy electron. In order to better understand this problem, the energy (pulse-height of the shaped signal) can be plot as a function of the baseline value in a 2-D histogram (Fig. \ref{TimeDiffOldDaq2D}). This analysis reveals a drift in the calculated energy for low baseline value. This phenomenon is probably due to an overcompensation of the amplification of the shaper for low baseline value: after high-energy events the shaper baseline can be shifted. A compensation of this effect has been developped within the shaper.

\paragraph{New DAQ}

\begin{figure}[h!]
\begin{center}
\includegraphics[width=100mm]{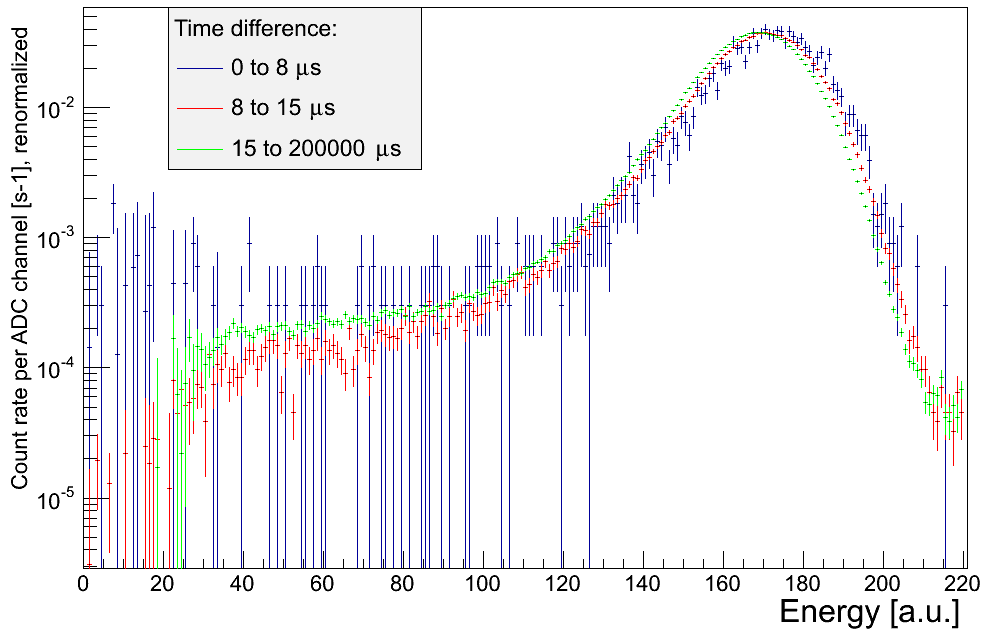}
\end{center}
\caption[Time difference spectrum after high-energy electron, channel 1, new DAQ.] {Time difference spectrum after high-energy electron for the channel 1 with the new DAQ.}
\label{timeDiff_ch1_NewDaq}
\end{figure}

The same analysis were made with the new acquisition system, as shown in Fig. \ref{timeDiff_ch1_NewDaq}. No problem detected, it is exactly the expected behavior. The very small energy difference between the three curves is related to the time of flight of protons inside the spectrometer. The more energy a proton have, the less time it will take for it to reach the detector.

\begin{figure}[h!]
\begin{center}
\includegraphics[width=100mm]{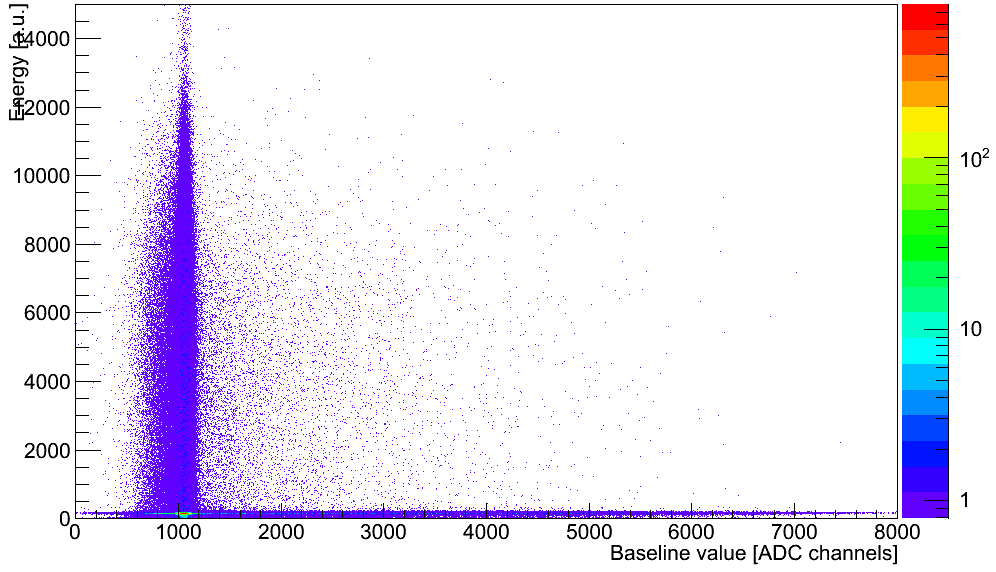}
\end{center}
\caption[Energy VS baseline value with the new DAQ, channel 1.] {Energy VS baseline value with the new DAQ, channel 1.}
\label{EvsBaseline_ch1}
\end{figure}

\begin{figure}[h!]
\begin{center}
\includegraphics[width=100mm]{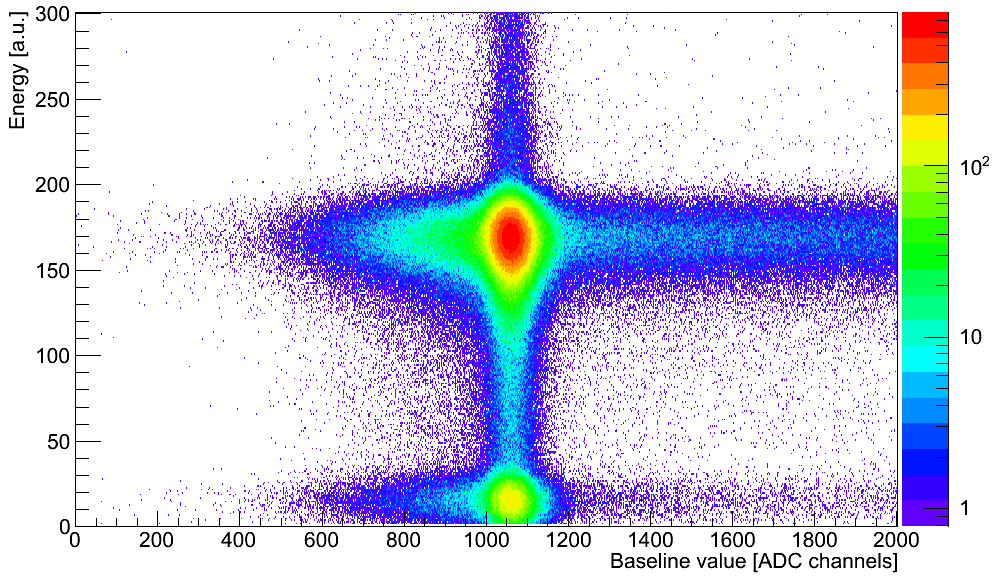}
\end{center}
\caption[Zoom of the energy VS baseline with the new DAQ, channel 1.] {Zoom of the energy VS baseline with the new DAQ, channel 1.}
\label{EvsBaseline_ch1_zoom2}
\end{figure}

The 2-D plot of the energy as a function of the baseline value is shown in Fig. \ref{EvsBaseline_ch1} and Fig. \ref{EvsBaseline_ch1_zoom2} (zoom). Because of the different pulse signals (shaped with the old DAQ and not with the new one) this plot has to be interpreted differently than Fig. \ref{TimeDiffOldDaq2D}.

On those two plots, we can see that the proton peak position is constant, whatever the baseline value is (horizontal line for Energy $\approx$ 180). There is no shift due to high-energy events and no apparent loss of proton events under certain circumstances.

\subsubsection{Saturation effect}

\begin{figure}[h!]
\begin{center}
\includegraphics[width=100mm]{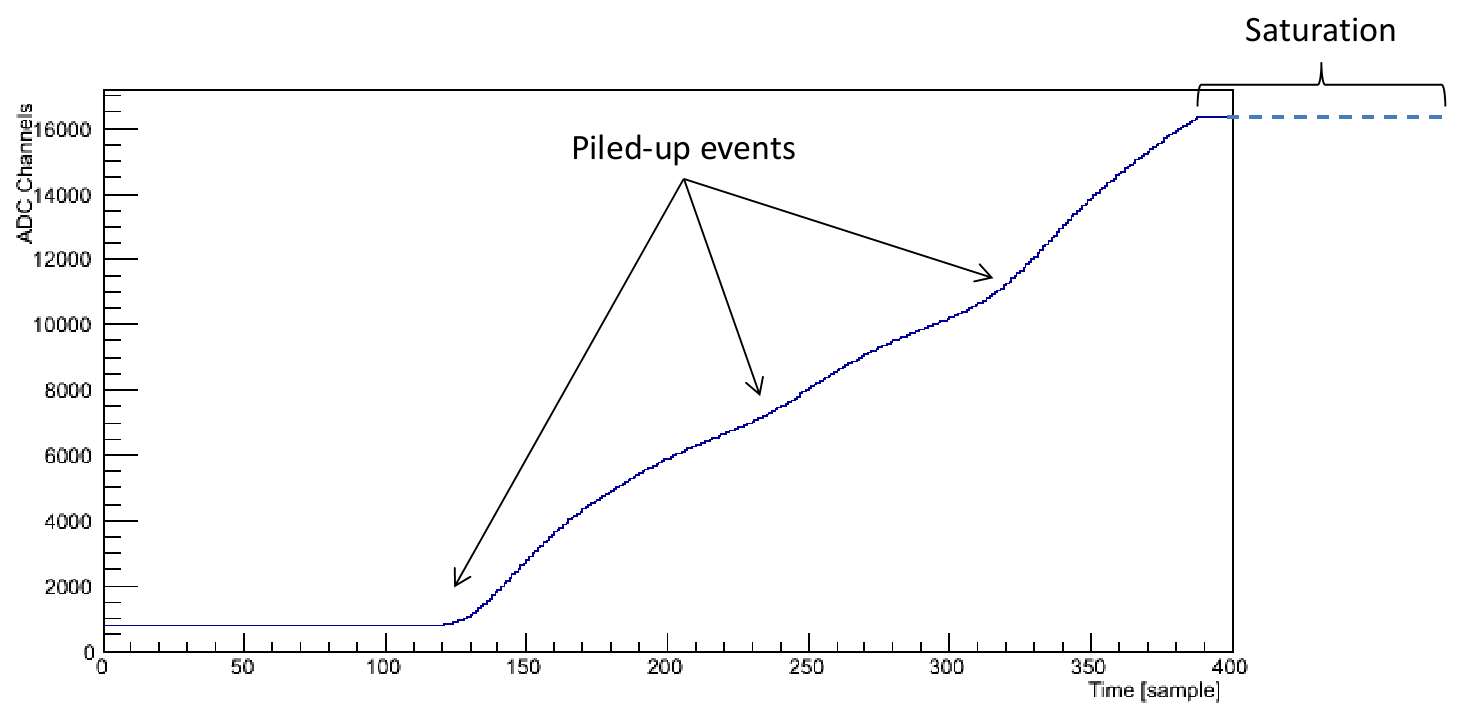}
\end{center}
\caption[Saturation of the ADC from piled-up events.] {Saturation of the ADC from high-energy piled-up events.}
\label{Saturation}
\end{figure}

When several high-energy events occurs very close to each other, the pile-up can be so intense that the concerned channel of the ADC can saturate (Fig. \ref{Saturation}). During the saturation, every event of this channel is lost and this duration depends on the energy of the responsible events. A too high saturation frequency can have a serious impact on the $a$ value, it is therefore important to quantify this effect.

\begin{figure}[h!]
\begin{center}
\includegraphics[width=80mm]{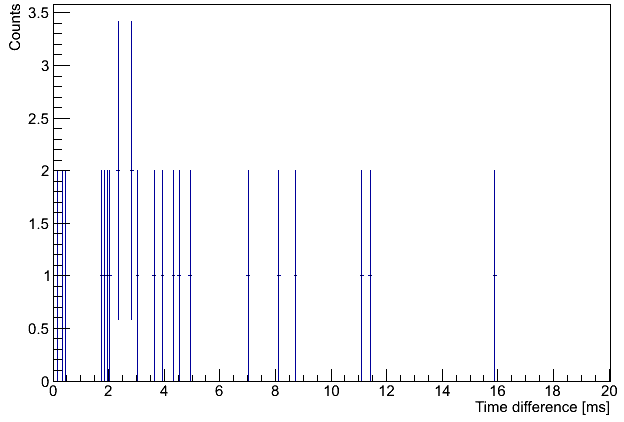}
\end{center}
\caption[Next event delay after saturation.] {Time difference between a saturation event and the next one on the same channel for the central pad. A data set of 72 files at AP=50V has been used.}
\label{saturation_ch1}
\end{figure}

Fig. \ref{saturation_ch1} represents the time difference between a saturating event (such as the one in Fig. \ref{Saturation}) and the next event on the same channel. The saturation duration can go up to $\approx15$ ms but no difference higher than $20$ ms have been measured. For a 72 files acquisition at AP=50 V (for a total of $8400$ sec), 23 saturations occured on channel 1, resulting in a $2.7\times 10^{-3}$ Hz saturation rate. In order to estimate this effect, the time difference mean value ($4.6$ ms for this channel and this data set) can be overestimated to $7$ ms to compensate for the low statistics. Since the maximum count rate in $a$ measurement conditions is $\approx440$ protons per second and per channel, the count rate for lost protons is: $7\times 10^{-3} \times 440 \times 2.7 \times 10^{-3} = 8.3\times 10 ^{-3}$ Hz. As a result, no more than one proton per channel is lost every 120 sec with the analysing plane at 50 V.

%%%%%%%%%%%%%%%%%%%%%%%%%%%%%%%%%%%%%%%%%%%%%%%%%%%%%%%%%%%%%%%%%%%%%%%%%%%%%%%
%  Conclusion
%%%%%%%%%%%%%%%%%%%%%%%%%%%%%%%%%%%%%%%%%%%%%%%%%%%%%%%%%%%%%%%%%%%%%%%%%%%%%%%
\clearpage
\section*{Conclusion} \addcontentsline{toc}{section}{Conclusion}
There were high expectations for the 2013 beam time of the $a$SPECT experiment. Despite the loss of several weeks of the precious beam time due to two consecutives bad detectors, the whole collaboration efforts allowed to take several $a$ measurement data sets and to implement all the past years upgrades. The new acquisition system has also been installed on the experiment but the lack of both (beam) time and experience with the new DAQ did not allow to acquire more than a few days with it without risking to loose too much time.

Nevertheless those measurements proved the good performance of this system for several key parameters for $a$SPECT, even if some results still need more statistics to be conclusive. This new acquisition system will be fully used for the next beam time of $a$SPECT.

The precise future of the $a$SPECT experiment is not yet decided but will be most certainly the measurement of another angular correlation coefficient: the proton asymmetry C. Another future addition to this experiment will be a tunable proton source, providing a 0 to 800 eV proton beam. It will be a great tool for the experiment, as it will allow to calibrate the detector and its electronics and will also permit to test the DAQ without any neutron beam.

Overall, this beam time has been for me a great experience. I have learn many different procedures and techniques related, for example, to the Ultra High Vacuum or to low temperatures. I had to work on different parts of the experiment, making it a very interesting work placement. It was also a great social experience, the whole collaboration greet me very nicely and I feeled truly integrated to the team. Even though the work rate was quite impressive, it was a very pleasant and internship, from both physics and personnal point of view.

%%%%%%%%%%%%%%%%%%%%%%%%%%%%%%%%%%%%%%%%%%%%%%%%%%%%%%%%%%%%%%%%%%%%%%%%%%%%%%%
%  Remerciements
%%%%%%%%%%%%%%%%%%%%%%%%%%%%%%%%%%%%%%%%%%%%%%%%%%%%%%%%%%%%%%%%%%%%%%%%%%%%%%%

\section*{Acknowledgements} \addcontentsline{toc}{section}{Acknowledgements}
I would like to especially thank my internship supervisor Torsten Soldner. He gave me the oppurtunity to participate in a great adventure, gave me responsibilies and trusted me. He also helped me whenever I was in need, always with very good explanations. Furthermore, I would like to thank Romain Maisonobe for his help, his kindness and also all the great laughs we had. Another person I would like to personnaly thank is Martin Simson. He introduced me to the $a$SPECT electronics and always took time to nicely fix whatever I broke.

I would also like to thanks all the $a$SPECT collaboration: Alexander Wunderle, Christian Schmidt, Gertrud Konrad, Oliver Zimmer, Marcus Beck, Stefan Baessler, and Werner Heil. Finally I would also like to thank all the NPP group for being such nice people: Felix, Christine, Damien, Camille, Florian, Aurelien, Angelika...

%%%%%%%%%%%%%%%%%%%%%%%%%%%%%%%%%%%%%%%%%%%%%%%%%%%%%%%%%%%%%%%%%%%%%%%%%%%%%%%
%  Biblio
%%%%%%%%%%%%%%%%%%%%%%%%%%%%%%%%%%%%%%%%%%%%%%%%%%%%%%%%%%%%%%%%%%%%%%%%%%%%%%%

\newpage
 \addcontentsline{toc}{section}{References}
\bibliographystyle{alpha} % Le style est mis entre accolades.
%\bibliography{BibliographyDB}

\newcommand{\etalchar}[1]{$^{#1}$}

% Fin du document

\clearpage

\textbf{Abstracts:}

\vspace{5cm}

L'experience aSPECT a pour but de mesurer le coefficient de correlation angulaire entre l'electron et l'antineutron \textit{a} lors de la desintegration $\beta$ du neutron libre. Ce coefficient peut etre retrouver en mesurant le spectre de recul des protons produits lors de la decroissance des neutrons. D'avril a aout 2013, $a$SPECT a ete installe sur la ligne de neutrons froids PF1b de l'Institut Laue-Langevin (ILL).

Mon stage a eu pour objectif de tester un nouveau systeme d'acquisition dernier-cri et d'en verifier le comportement dans le cadre de cette experience de haute-precision.

\vspace{5cm}

From April 2013 to the beginning of August 2013, the aSPECT spectrometer was installed on a cold neutron beam facility at the Institut Laue-Langevin (ILL) in order to realize a high-precision measurement of a. Among all the upgrades that were made during the past years, a new state-of-the-art acquisition system has been tested during this beam time.

My internship consisted of testing this new acquisition system and to check his good behavior on this high-precision experiment.

\end{document}